\documentclass[aps, prd, twocolumn, amsmath, floats,floatfix, superscriptaddress, nofootinbib]{revtex4}

\usepackage{amssymb}
\usepackage{amsmath}
\usepackage{verbatim}
\usepackage{mathrsfs}
\usepackage{amsfonts}
\usepackage{latexsym}
\usepackage{epsfig}
\usepackage{color}
\usepackage{graphicx,subfigure}
\usepackage{units}
\usepackage{envmath}
\usepackage{natbib}
\usepackage{ctable}
\usepackage{soul}
\usepackage{ulem}
\begin{document}

%%%%%%%%%%%%%%%%%%
%%%   MACROS   %%%
%%%%%%%%%%%%%%%%%%

\definecolor{orange}{rgb}{0.9,0.45,0}

\newcommand{\re}{\mbox{Re}}
\newcommand{\im}{\mbox{Im}}
\newcommand{\tf}[1]{\textcolor{red}{TF: #1}}
\newcommand{\nsg}[1]{\textcolor{cyan}{#1}}
\newcommand{\saeed}[1]{\textcolor{blue}{SF: #1}}
\newcommand{\fdg}[1]{\textcolor{orange}{FDG: #1}}
\newcommand{\jc}[1]{\textcolor{magenta}{JC: #1}}

\def\CovDev{D}
\def\Res{{\mathcal R}}
\def\Gammaflat{\hat \Gamma}
\def\metricflat{\hat \gamma}
\def\Dflat{\hat {\mathcal D}}
\def\part_n{\partial_\perp}

%=== Definition for abbreviations ===
\def\Lie{\mathcal{L}}
\def\A{\mathcal{X}}
\def\Aphi{\A_{\phi}}
\def\hAphi{\hat{\A}_{\phi}}
\def\E{\mathcal{E}}
\def\Ham{\mathcal{H}}
\def\M{\mathcal{M}}
\def\R{\mathcal{R}}
\def\p{\partial}

\def\hg{\hat{\gamma}}
\def\hA{\hat{A}}
\def\hD{\hat{D}}
\def\hE{\hat{E}}
\def\hR{\hat{R}}
\def\hcA{\hat{\mathcal{A}}}
\def\hDelt{\hat{\triangle}}

\def\na{\nabla}
\def\dif{{\rm{d}}}
\def\non{\nonumber}
\newcommand{\erf}{\textrm{erf}}
%====================================

\renewcommand{\t}{\times}

\long\def\symbolfootnote[#1]#2{\begingroup%
\def\thefootnote{\fnsymbol{footnote}}\footnote[#1]{#2}\endgroup}

%%%%%%%%%%%%%%%%%
%%%   TITLE   %%%
%%%%%%%%%%%%%%%%%

\title{A stabilization mechanism for excited fermion-boson stars}

\author{Fabrizio Di Giovanni}
\affiliation{Departamento de
  Astronom\'{\i}a y Astrof\'{\i}sica, Universitat de Val\`encia,
  Dr. Moliner 50, 46100, Burjassot (Val\`encia), Spain}
  
\author{Saeed Fakhry}
\affiliation{Department of Physics, Shahid Beheshti University, G. C., Evin, Tehran 19839, Iran}
\affiliation{Departamento de Astronom\'{\i}a y Astrof\'{\i}sica, Universitat de Val\`encia, Dr. Moliner 50, 46100, Burjassot (Val\`encia), Spain}

\author{Nicolas Sanchis-Gual}
\affiliation{Departamento  de  Matem\'atica  da  Universidade  de  Aveiro  and  Centre  for  Research  and  Developmentin  Mathematics  and  Applications  (CIDMA),  Campus  de  Santiago,  3810-183  Aveiro,  Portugal}

\author{Juan Carlos Degollado} 
\affiliation{Instituto de Ciencias F\'isicas, Universidad Nacional Aut\'onoma 
de M\'exico, Apdo. Postal 48-3, 62251, Cuernavaca, Morelos, M\'exico}

\author{Jos\'e A. Font}
\affiliation{Departamento de
  Astronom\'{\i}a y Astrof\'{\i}sica, Universitat de Val\`encia,
  Dr. Moliner 50, 46100, Burjassot (Val\`encia), Spain}
\affiliation{Observatori Astron\`omic, Universitat de Val\`encia, C/ Catedr\'atico 
  Jos\'e Beltr\'an 2, 46980, Paterna (Val\`encia), Spain}

\begin{abstract} 
We study numerically the nonlinear stability of {\it excited} fermion-boson 
stars in spherical symmetry. Such compound hypothetical stars, composed by 
fermions and bosons, are gravitationally bound, regular, and static 
configurations described within the coupled Einstein-Klein-Gordon-Euler 
theoretical framework. The excited configurations are characterized by the 
presence in the radial profile of the (complex, massive) scalar field -- the 
bosonic piece -- of at least one node across the star. The dynamical emergence 
of one such configuration from the accretion of a cloud of scalar field onto an 
already-formed neutron star, was numerically revealed in our previous 
investigation. Prompted by that finding we construct here equilibrium 
configurations of excited fermion-boson stars and  study their stability 
properties using numerical-relativity simulations. In addition, we also analyze 
their dynamical formation from generic, constraint-satisfying initial data.  
Contrary to purely boson stars in the excited state, which are known to be 
generically unstable, our study reveals the appearance of a cooperative 
stabilization mechanism between the fermionic and bosonic constituents of those 
excited-state mixed stars. While similar examples of stabilization mechanisms 
have been recently discussed in the context of $\ell-$boson stars and 
multi-field, multi- frequency boson stars, our results seem to indicate that the 
stabilization mechanism is a purely gravitational effect and does not depend on 
the type of matter of the companion star.
\end{abstract}

\maketitle

\vspace{0.0cm}

%%%%%%%%%%%%%%%%%%%%%%
\section{Introduction}
%%%%%%%%%%%%%%%%%%%%%%

The nature of Dark Matter (DM) is an outstanding open issue in modern cosmology. Abundant evidence in support of its existence has been collected, starting with observations of galaxy rotation curves, gravitational lensing, and the cosmic microwave background~\cite{Hinshaw:2012aka,Reid:2012sw,Hu:2000ke,Caldwell:2009ix,Blake:2011en,Hlozek:2014lca,Chluba:2019nxa,Abazajian:2019eic}. 
Since those indications arise only through gravitational effects, gravitational interactions are a promising channel to unveil the nature of DM.  Although several possibilities  have been proposed, it has been recognized that ultralight boson fields with masses of the order of $10^{-22}$ eV are a compelling candidate as the main component of DM 
~\cite{Liddle:1993ha,Hu:1998kj,Matos:1999et,Matos:2000ng,
Amendola:2005ad,Arvanitaki:2009fg,Lundgren:2010sp,
Marsh:2010wq,Hui:2016ltb}. Bosons can clump together to form self-gravitating equilibrium states, known as boson stars, which provides a natural alternative to standard structure formation through DM seeds (see~\cite{Jetzer:1991jr,Schunck:2003kk,Liebling:2012fv}).
 
Kaup~\cite{Kaup:1968zz} and Ruffini and Bonazzola~\cite{Ruffini:1969qy} pioneered the investigation of boson stars. Their studies showed that the mass of a boson star is $\sim M^2_{\rm Pl}/\mu$ and that its characteristic size is of the order of the  Compton wavelength of the boson particle, $\sim 1/\mu$, where $\mu$ is the mass of the particle and $M_{\rm Pl}$ is the Planck mass. Later, Colpi, Shapiro and Wasserman~\cite{Colpi:1986ye} incorporated self-interacting scalar particles and found that the corresponding boson stars have (larger) masses of $\sim \Lambda^{1/2}M^2_{\rm Pl}/\mu$, where $\Lambda$ is a parameter characterizing the strength of the self interaction. For $\Lambda^{1/2}\gg1$, this scaling is no longer valid and the maximum mass of the star turns out to be of the order of the Chandrasekhar mass $\sim M^3_{\rm Pl}/\mu^2$ for fermion stars~\cite{Jetzer:1991jr}.

The stability of equilibrium models of ground-state, spherical boson stars 
subject to perturbations has been studied  using both linear perturbation 
analysis~\cite{Gleiser:1988rq,Lee:1988av} and nonlinear numerical 
simulations~\cite{Balakrishna:1997ej, Seidel:1990jh}. This body of work has 
showed that  ground-state models -- the so-called fundamental family -- are 
stable  as long as the central value of the scalar field, $\phi_c$, is smaller 
than that of the configuration with the maximum Arnowitt-Desser-Misner (ADM) 
mass. These findings support the hypothesis that boson stars may form 
dynamically  under general initial conditions, as shown 
by~\cite{seidel1994formation}. The stability of {\it excited} boson stars, 
i.e.~stars for which the scalar-field amplitude exhibits at least one radial 
node across the star, was investigated in~\cite{Lee:1988av,Balakrishna:1997ej}. 
Equilibrium configurations were also found for excited-state models. However, 
those are intrinsically unstable under generic perturbations: the excited-state 
configurations decay to the ground-state, collapse to a black hole, or disperse 
away. 

As already noticed in~\cite{Ruffini:1969qy} there exists the possibility that 
bosons within a boson star are not all in the ground state, but rather 
populating different coexisting states forming multi-state boson stars. In 
Ref.~\cite{Bernal:2009zy} Bernal \textit{et al} studied the dynamical evolution 
of perturbed multi-state boson stars demonstrating that stable states can form 
when the number of particles in the first excited state is smaller than the 
number of particles in the ground state. As we shall see, this type of 
stabilization mechanism has in part motivated the study we present here.

In a recent work~\cite{DiGiovanni2020a} we investigated the properties of 
macroscopic astrophysical objects that contain both bosons and fermions, known 
as fermion-boson stars~\cite{HENRIQUES1990511,LOPES199280}. The study  of  the  
dynamics  of  these compound  objects  is  important for  a  number  of  
reasons, ranging from the way they interact with surrounding matter to their 
stability~\cite{PhysRevD.87.084040,PhysRevD.102.064038}. 
In~\cite{DiGiovanni2020a} we built spherically-symmetric equilibrium 
configurations of fermion-boson stars and studied their nonlinear dynamical 
stability, through numerical-relativity simulations, under generic radial 
perturbations. Moreover,  we presented a dynamical scenario in which 
fermion-boson stars might form following the gravitational collapse of a cloud 
of scalar field surrounding an already-formed neutron star.  The equilibrium 
models considered in~\cite{DiGiovanni2020a} were all ground-state solutions, 
where the scalar field has no radial nodes across the star. However, the 
simulations of~\cite{DiGiovanni2020a} that dealt with the dynamical formation 
scenario revealed the emergence of a final configuration with a node in the 
scalar field, similar to a static solution with an excited state. This suggests 
that there might be {\it stable} fermion-boson stars with nodes. 

To investigate this issue we shall consider here spherically-symmetric 
equilibrium solutions of excited-state fermion-boson stars, i.e.~configurations 
in which the radial profile of the boson part has at least one radial node 
across the star, analizing their nonlinear dynamics. Both, models with and 
without self-interaction in the bosonic part will be considered. In addition,  
the formation scenario will receive extra attention here, by discussing new 
evolutions of neutron stars surrounded by scalar field clouds. Our investigation 
reveals the existence of a cooperative stabilization mechanism at work: the 
presence of a stable fermionic core that only interacts gravitationally with the 
scalar field stabilizes the excited state of the bosonic part of the compound 
star. We note that for purely boson stars in the excited state such mechanism is 
not active and those models are generically 
unstable~\cite{Lee:1988av,Balakrishna:1997ej}. Similar examples of 
stabilization mechanisms have been recently discussed in the context of 
$\ell-$boson stars~\cite{Alcubierre:2018ahf} and multi-field, multi-frequency 
boson stars~\cite{sanchis2021multi} (see 
also~\cite{Guzman:2019gqc,guzman2021stability}). While those studies have 
uncovered a cooperative stabilization mechanism by accounting for a second boson 
star (or a third one for $\ell-$boson stars) the results reported in this paper 
seem to indicate that the mechanism is a purely gravitational effect and does 
not depend on the type of matter of the companion star but rather on its 
dynamical properties. This effect could be similar to the {\it 
stabilization} of flat galactic rotation curves by dark matter halos in 
galaxies~\cite{rubin1978extended}.

This paper is organized as follows: In Section~\ref{sec2} we introduce the 
theoretical framework for mixed fermion-boson stars we will use to build the 
equilibrium configurations and we also introduce the corresponding evolution 
equations that will be used for the simulations. The actual equilibrium models 
are discussed in Section~\ref{sec:ini} which also describes the scenario for the 
dynamical formation of compound stars. In Section~\ref{sec:numerics} a brief 
description of the numerical framework is given. We do not go into much detail 
on purpose, since the framework is identical to that employed in our previous  
investigation~\cite{DiGiovanni2020a}. Section~\ref{results} presents our main 
results. Finally, the main conclusions of this work are reported  in 
Section~\ref{sec:conclusions}. Throughout the manuscript Greek indices are 
spacetime while Latin indices are purely spatial. For our simulations we set $G 
= c = \hbar = 1$, where $G$ is Newton's gravitational constant, $c$ is the 
speed of light and $\hbar$ is the reduced Planck's constant. 

%%%%%%%%%%%%%%%%%%%%%%
\section{Setup}
\label{sec2}
%%%%%%%%%%%%%%%%%%%%%%
%%%%%%%%%%%%%%%%%%%%%%

Our setup is the same one as in our recent work~\cite{DiGiovanni2020a}. Therefore, here we avoid unnecessary repetition and we focus on the basic equations that are needed in the definitions of physical quantities that will be used throughout the paper. The interested reader is addressed to~\cite{DiGiovanni2020a} for further details. 

\subsection{Matter models} \label{matter}

We study models of fermion-boson stars in which the bosonic matter and the fermionic matter only interact through gravity. Therefore, the total stress-energy tensor describing the physical system is the sum of two contributions, one from a complex scalar field and one from a perfect fluid:
\begin{eqnarray}
T_{\mu\nu}&=& T_{\mu\nu}^{\rm{fluid}} + T_{\mu\nu}^{\phi} ,
\end{eqnarray}
where
\begin{eqnarray}
T_{\mu\nu}^{\rm{fluid}}&=& [\rho(1+\epsilon) + P] u_{\mu}u_{\nu} + P g_{\mu\nu}, \\
T_{\mu\nu}^{\phi}&=& - \frac{1}{2}g_{\mu\nu}\partial_{\alpha}\bar{\phi}\partial^{\alpha}\phi - V(\phi) \nonumber \\
 &+& \frac{1}{2}(\partial_{\mu}\bar{\phi}\partial_{\nu}\phi+\partial_{\mu}\phi\partial_{\nu}\bar{\phi}) .
\end{eqnarray}
The fermionic matter is described by the fluid pressure $P$, its rest-mass density $\rho$, its internal energy $\epsilon$ and its 4-velocity $u^{\mu}$. The scalar-field potential is defined as
\begin{eqnarray} \label{s0}
V(\phi) =  \frac{1}{2} \mu^2\bar{\phi}\phi+\frac{1}{4}\lambda(\bar{\phi}\phi)^2,
\end{eqnarray}
where $\mu$ and $\lambda$ are the mass and the self-interaction parameter of the bosonic particle, respectively. The bar symbol denotes complex conjugation. The equations of motion are given by the conservation laws of the stress-energy tensor and of the baryonic particles for the perfect fluid, and by the Klein-Gordon equation for the complex scalar field, together with the Einstein equations for the spacetime dynamics. The system is then closed by an equation of state (EoS) for the fluid. For simplicity we choose both a (zero-temperature) polytropic EoS and an ideal-gas EoS,
\begin{equation}\label{EOS}
P= K \rho^{\Gamma} = (\Gamma-1)\rho\epsilon\,,
\end{equation}
where $K$ is the polytropic constant and $\Gamma$ the adiabatic index. We employ the polytropic EoS to construct the equilibrium configurations while the evolution code implements the $\Gamma$-law equation as it allows to take into account potential shock-heating effects during the simulations. All equilibrium models are built using $K=100$ and $\Gamma = 2$. 

\subsection{Equilibrium configuration equations} \label{stat}

In order to construct the equilibrium configurations we assume a static and spherically-symmetric metric in Schwarzschild coordinates 
\begin{eqnarray} \label{Schwarzschild_metric}
ds^2 = -\alpha(r)^2 dt^2 + \tilde{a}(r)^2 dr^2 + r^2 ( d\theta^2 + \sin{\theta}^2 d\varphi^2),
\end{eqnarray}
written in terms of two geometrical functions $\tilde{a}(r)$ and $\alpha(r)$.

The boson star is described by a harmonic time dependence for the complex scalar field, $\phi(t, r) = \phi(r) e^{-i\omega t}$, where $\omega$ is its eigenfrequency. We employ a quartic self-interaction potential as defined in Eq.~\eqref{s0}, where we replace the self-interaction parameter $\lambda$ by the dimensionless variable
\begin{eqnarray}
\Lambda=\frac{M_{\rm Pl}^{2}\lambda}{4\pi \mu^{2}},
\end{eqnarray}
where $M_{\rm Pl} = \sqrt{\hbar c/G}$ is the Planck mass (which is one in our units). As in~\cite{DiGiovanni2020a} we use the mass of the boson particle to rescale the radial coordinate, the mass of the star, the time, and the frequency according to $r\rightarrow r\mu$, $M\rightarrow M\mu$,  $t\rightarrow t\mu$, and $\omega\rightarrow \omega/\mu$. Details on this scaling and on how to recover the physical units from those used in the numerical code are provided in~\cite{DiGiovanni2020a}. In the same reference the interested reader can find the set of ordinary differential equations (ODEs) that we solve to obtain the equilibrium configurations.

\subsection{Evolution equations} \label{sec:basic}

The formalism of the numerical evolutions relies on a spherically-symmetric metric in isotropic coordinates
\begin{eqnarray}\label{isotropic_metric}
ds^2 = -\alpha(\hat{r})^2 dt^2 + \psi(\hat{r})^4 \gamma_{ij} (dx^{i} + \beta^{i}dt)(dx^{j} + \beta^{j}dt),
\end{eqnarray} 
where $\alpha$ is the lapse function, $\beta^{i}$ is the shift vector, and $\psi(\hat{r})$ is a conformal factor. The spatial 3-metric components are
\begin{eqnarray}
\gamma_{ij} dx^idx^j = a(\hat{r})d\hat{r}^2 + b(\hat{r})\hat{r}^2 (d\theta^2 + \sin{\theta}^2 d\varphi^2) \,.
\end{eqnarray}
Note that $a$ and $\tilde{a}$ should not be confused, as they refer to two different metrics; the hat symbol is used to distinguish the isotropic radial coordinate from the Schwarzschild one. From now on, to simplify the notation, we will neglect the hat in the radial coordinate, keeping in mind that $r$ will refer to the isotropic radial coordinate. 

We follow Brown's covariant form~\cite{Brown:2009,Alcubierre:2010is} of the 
Baumgarte-Shapiro-Shibata-Nakamura (BSSN) formulation of Einstein's 
equations~\cite{nakamura1987general,Shibata95, Baumgarte98} to perform our 
numerical evolutions. The evolved quantities are the spatial metric 
$\gamma_{ij}$, the BSSN conformal factor $\chi$ , the trace of the extrinsic 
curvature $K$, its traceless part $A_a = A^r_r$, 
$A_{b}=A^{\theta}_{\theta}=A^{\varphi}_{\varphi}$, and the radial component of 
the BSSN conformal connection functions $\Delta^r$. The reader is addressed 
to Ref.~\cite{Shibata95, Baumgarte98} for definitions of those quantities and 
to Ref.~\cite{Montero:2012yr} for details of the full system of evolution 
equations we solve and on the gauge conditions..

The matter source terms appearing in the evolution equations arise from projections of the total stress-energy tensor $T_{\mu\nu}$. Those are the energy density $\mathcal{E}$, the momentum density $j_{i}$ measured by a normal observer $n^{\mu}$, and the spatial projection of the energy-momentum tensor $S_{ij}$, and read:
\begin{align}\label{matter_source_terms}
\mathcal{E}&= n^{\mu}n^{\nu}T_{\mu \nu}, \\
j_i&=-\gamma_{i}^{\mu}n^{\nu}T_{\mu \nu}, \\
S_{ij}&= \gamma_{i}^{\mu} \gamma_{j}^{\nu} T_{\mu \nu}.
\end{align}
These quantities are obtained for both the fluid and the scalar field, considering $T_{\mu\nu}^{\rm{fluid}}$ or $T_{\mu\nu}^{\phi}$, respectively. Again, explicit expressions of the matter source terms and of our first-order system of evolution and constraint equations are reported in~\cite{DiGiovanni2020a}.

%%%%%%%%%%%%%%%%%%%%%%
\section{Initial Data} \label{sec:ini}
%%%%%%%%%%%%%%%%%%%%%%
%%%%%%%%%%%%%%%%%%%%%%

\subsection{Equilibrium configurations} \label{sec:initial_static}

We solve the set of ODEs alluded to in Section~\ref{stat} (see~\cite{DiGiovanni2020a} for details) to construct suitable initial data representing equilibrium configurations of fermion-boson stars. The system of ODEs is written as an eigenvalue problem for the frequency of the scalar field $\omega$, which depends on two parameters, the central values of the scalar field $\phi_{c}$ and of the fermionic rest-mass density $\rho_{c}$. We adopt the two-parameter shooting method to find the eigenfrequency $\omega_{\rm shoot}$ corresponding to an excited state of the scalar field. Once $\omega_{\rm shoot}$ is found, we use a 4th-order Runge-Kutta method to integrate the ODEs and reconstruct the entire solution. Finally we rescale both the lapse function $\alpha$ and $\omega_{\rm shoot}$ to impose Schwarzschild outer boundary conditions. We require regularity at the origin to be satisfied by the metric functions, together with a vanishing scalar field at the outer boundary. Hence, the boundary conditions read as follows:
\begin{eqnarray}
&\tilde{a}(0) = 1, \hspace{0.3cm} & \phi(0) = \phi_{c}, \nonumber\\
&\alpha(0) = 1, \hspace{0.3cm} &  \lim_{r\rightarrow\infty}\alpha(r)=\lim_{r\rightarrow\infty}\frac{1}{\tilde{a}(r)},\nonumber\\
& \Psi(0)=0, \hspace{0.3cm} & \lim_{r\rightarrow\infty}\phi(r)=0, \nonumber\\
&\rho(0) = \rho_{c},  & \hspace{0.3cm}  P(0)=K\rho_{c}^{\Gamma},  \hspace{0.3cm} \lim_{r\rightarrow\infty}P(r)=0.
\end{eqnarray}

Purely boson-star models can be built solving the set of ODEs assuming $\rho_c=0$. For such stars it is known~\cite{Kaup:1968zz,Ruffini:1969qy} that there is a countably infinite set of solutions, labelled by the number of nodes in the radial profile of the scalar field, $n$. Nodeless solutions, $n=0$, are considered to be the ground-state solutions, while all other $n\neq 0$ solutions are excited states. 

We next define some useful physical quantities that describe the properties of the equilibrium configurations. The total gravitational mass can be computed from the value of the metric coefficients at infinity, and reads
\begin{eqnarray} \label{mass}
M_{\rm T}=\lim_{r\longrightarrow\infty}\frac{r}{2}\left(1-\frac{1}{\tilde{a}^2}\right),
\end{eqnarray}
which coincides with the ADM mass at infinity. Noether's theorem predicts the 
existence of a conserved charge related to the invariance of the Klein-Gordon 
Lagrangian under global U(1) transformations of the scalar field, $\phi 
\rightarrow \phi\,e^{i\delta}$. This charge is associated with the number of 
bosonic particles $N_{\rm B}$. Moreover, a definition of the number of fermionic 
particles $N_{\rm F}$ follows by the conservation of the baryonic number. These 
conserved charges can be evaluated by integrating their volume densities as 
follows:
\begin{eqnarray}
N_{\rm B} = 4 \pi \int \frac{\tilde{a} \omega \phi^2 r^2}{\alpha} \,dr, \hspace*{0.5cm} N_{\rm F}= 4 \pi \int \tilde{a} \rho r^2 \,dr.
\end{eqnarray}
Finally, we define the radius of the bosonic (fermionic) contribution to the fermion-boson star, $R_{\rm B}$($R_{\rm F}$), as the radius of the sphere containing $99\%$ of the corresponding particles.

\begin{figure}[t!]
\begin{minipage}{1\linewidth}
\includegraphics[width=1.0\textwidth]{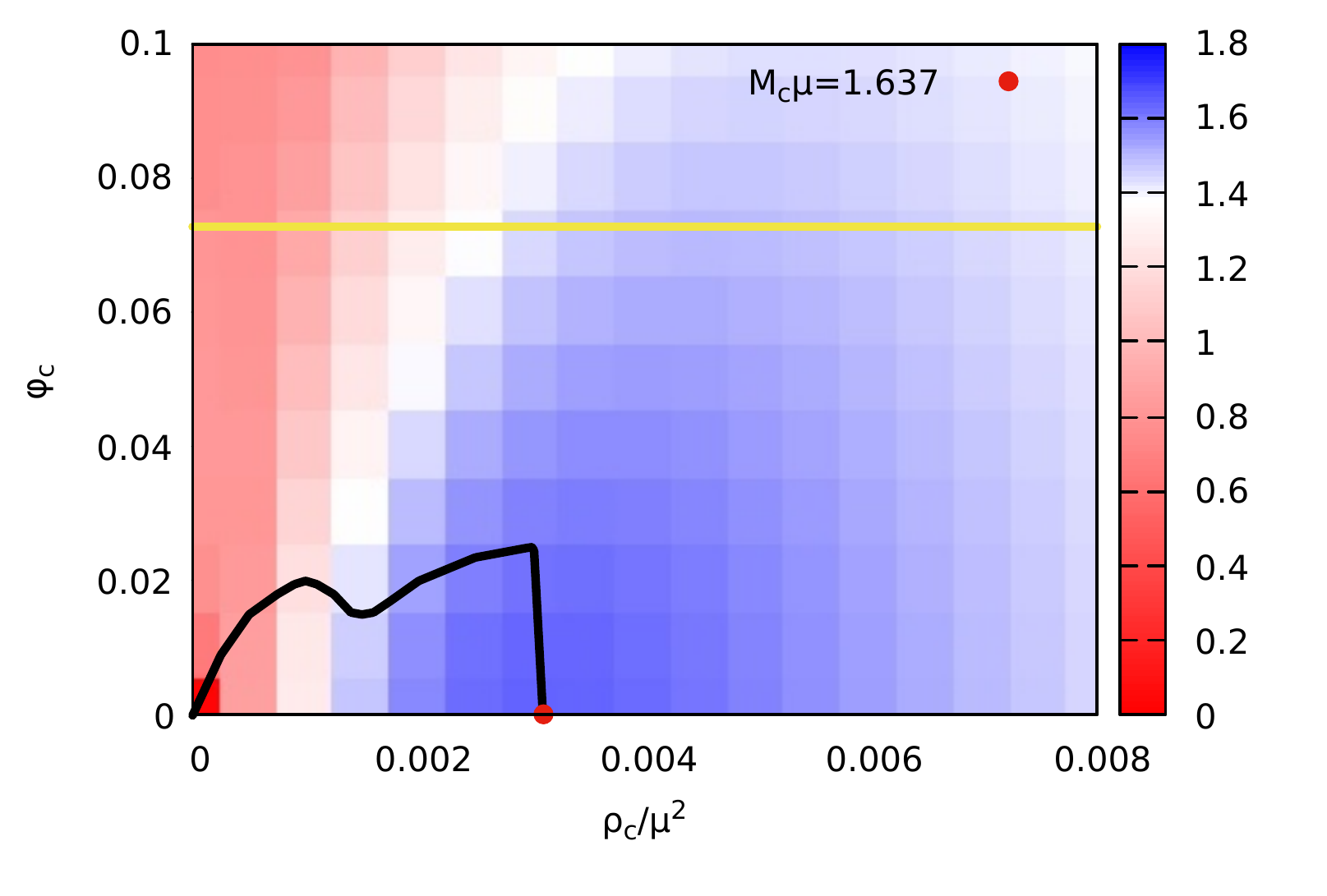}  
\includegraphics[width=1.0\textwidth]{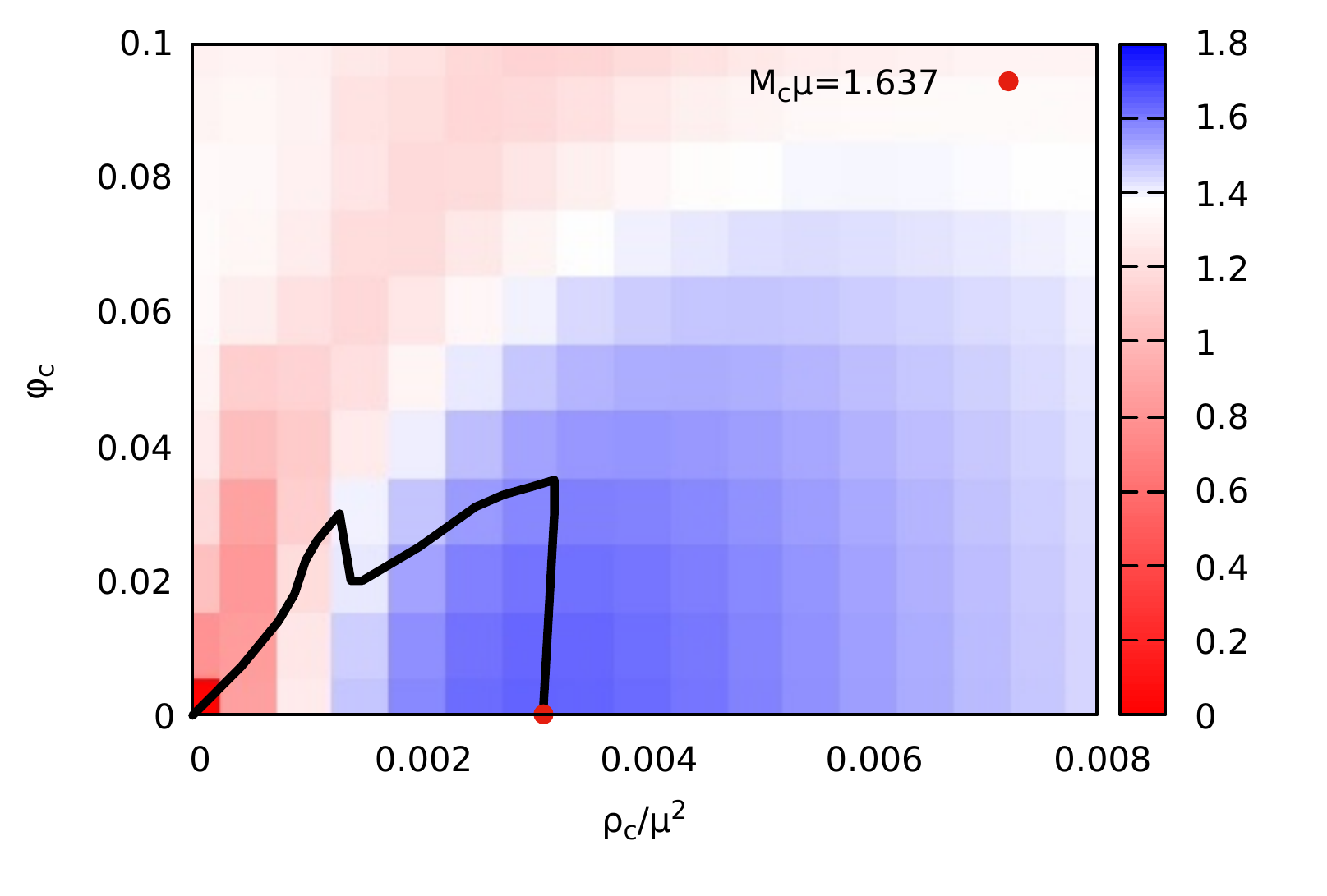} 
\includegraphics[width=1.0\textwidth]{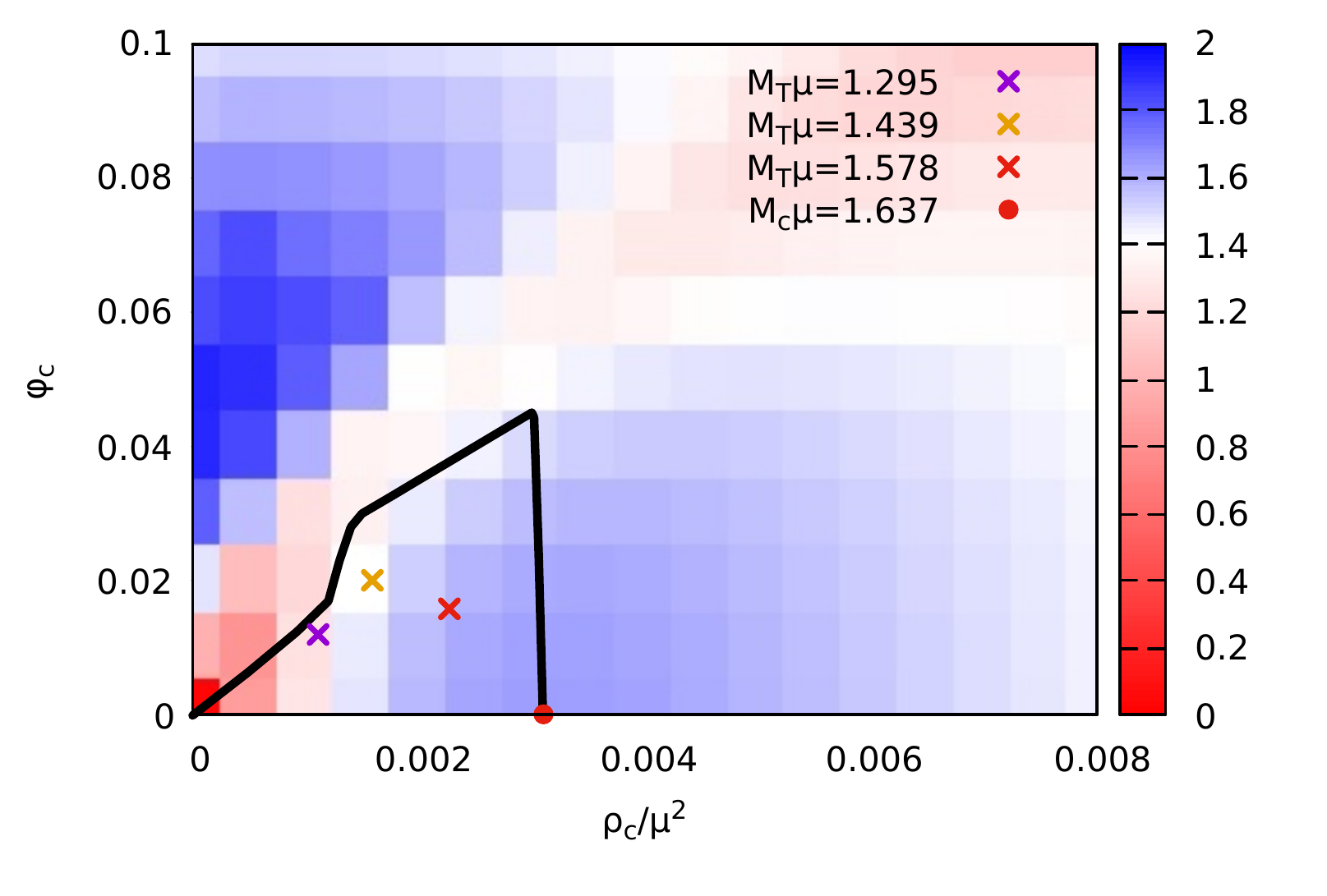} 
\caption{Equilibrium configurations of $n=1$ excited-state, fermion-boson stars for $\Lambda = -30$ (top), $\Lambda = 0$ (middle), and $\Lambda = 30$ (bottom). The black solid lines depict the boundary between stable models (bottom-left-corner regions of the plots)  and unstable models. The yellow solid line for the case $\Lambda = -30$ indicates the maximum value of $\phi_c$ that ensures the non-negativity of the scalar field potential $V(\phi)$ in the entire spatial domain. } 
\label{fig:existence_plot} 
\end{minipage}
\end{figure}

In Fig.~\ref{fig:existence_plot} we depict the mass of the $n=1$ mixed star models, Eq.~\eqref{mass}, as a function of the two parameters, $\rho_c$ and $\phi_c$, for three different values of $\Lambda$. As fermion stars do not depend on $\Lambda$ their threshold (or critical) mass, $M_c\mu$, which is highlighted with red dots on the $x$-axis of all plots in Fig.~\ref{fig:existence_plot}, is constant for all values of $\Lambda$ and equal to $M_c\mu=1.637$. The black solid lines in the different panels of Fig.~\ref{fig:existence_plot} indicate the boundary separating stable and unstable regions  in the parameter space. A comparison with the corresponding existence plot of $(n=0)$ ground-state mixed-star solutions (see Fig.~1 of~\cite{DiGiovanni2020a}) shows that the stability region shrinks significantly for the $n=1$ excited-state models.

\begin{figure}[t!]
\begin{minipage}{1\linewidth}
\includegraphics[width=1.0\textwidth]{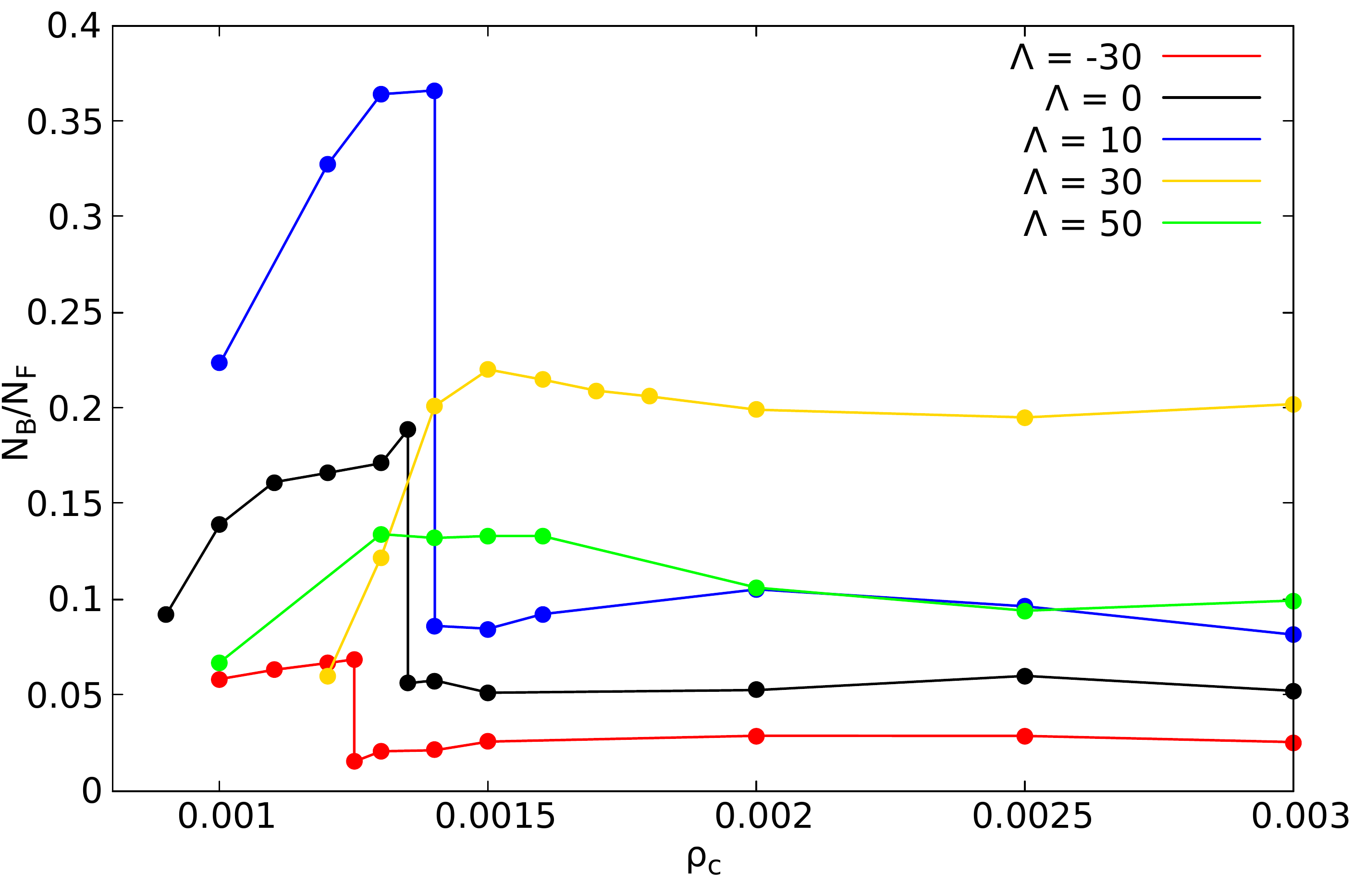}  
\caption{Ratio between bosonic and fermionic particles, $N_{\rm B}/N_{\rm F}$, as a function of the central density of fermionic matter, $\rho_{c}$, for $\Lambda =-30, 0, 10, 30, 50$. The ratio is computed for mixed-star models  at the boundary between stable and unstable regions (indicated by the filled circles).} 
\label{fig:NbNf_l-30_l0_l30} 
\end{minipage}
\end{figure}

To assess the stability of excited fermion-boson stars (and to be able to draw the black lines in the figure) we rely on nonlinear numerical evolutions. While it might be interesting to carry out a linear stability analysis of the models, it is unclear if such an analysis would provide meaningful information, telling from earlier results with excited boson stars~\cite{Lee:1988av,Balakrishna:1997ej}. To construct the black lines in Fig.~\ref{fig:existence_plot} we explore the region of the parameter space close to the stable branch of the neutron star models (which is the $x$-axis in the plots) because we expect to see stability only if the contribution of the fermionic part is large enough to stabilise an otherwise unstable excited scalar field. By performing numerical evolutions of models in this region, keeping fixed the value of $\rho_c$ and increasing $\phi_c$, it is possible to  find the first unstable model. As expected, we find a region in the parameter space where excited fermion-boson stars are stable, for the three values of the self-interaction parameter $\Lambda$ considered. The red crosses in the $\Lambda=30$ plot (bottom panel in Fig.~\ref{fig:existence_plot}) are models that we could also form dynamically and which we discuss in detail below. As we pointed out in our previous work~\cite{DiGiovanni2020a} considering negative values of $\Lambda$ raises the issue that the scalar potential may break the weak-energy condition (see e.g.~the discussion in~\cite{Barcelo2000}). The horizontal yellow line in the plot for $\Lambda =-30$ (top panel) indicates the maximum central value of $\phi$ that ensures the non-negativity of the scalar-field potential, which is $\phi_{c}=0.0728$. We do not consider models above this line as their evolution might give rise to naked singularities.

 In their work on multi-state boson stars~\cite{Bernal:2009zy} Bernal {\it et al} found that models for which the number of bosonic particles in the ground state is higher than in the first excited state are stable, and unstable otherwise. A similar relation might occur between the number of fermionic and bosonic particles in the case of compound stars. Fig.~\ref{fig:NbNf_l-30_l0_l30} depicts the ratio between the number of bosonic and fermionic particles, $N_{\rm B}/N_{\rm F}$, as a function of $\rho_c$ for $\Lambda= -30$, 0, 10, 30, and 50.  
 We only consider models at the boundary between stable and unstable regions (the black solid lines in Fig.~\ref{fig:existence_plot}). We observe two different regimes. Up to a certain threshold value of the fluid central density, the value of $N_{\rm B}/N_{\rm F}$ increases monotonically. When a critical value $\rho_c^{\rm crit}$ is reached $N_{\rm B}/N_{\rm F}$ saturates and becomes roughly constant with $\rho_c$. The value of $\rho_c^{\rm crit}$ increases with $\Lambda$. Specifically we obtain $\rho_c^{\rm crit} = 0.00125, 0.00135, 0.0014, 0.0015, 0.0016$ for values of $\Lambda=-30, 0, 10, 30, 50$, respectively. We also observe that when the threshold value $\rho_c^{\rm crit}$ is reached, the ratio $N_{\rm B}/N_{\rm F}$ at the boundary between stable and unstable regions sharply drops, as shown in Fig.~\ref{fig:NbNf_l-30_l0_l30}. This drop disappears when the self-interaction parameter $\Lambda$ is high enough, somewhere between $10$ and $30$, as above $\Lambda=30$ the drop is not visible. 
 While the analogy with the findings of~\cite{Bernal:2009zy} for multi-state boson stars is not fully apparent for our models of excited-state mixed stars, we can nevertheless point out that having $N_{\rm B}/N_{\rm F}<1$ appears as a necessary condition for the stability of the models.

\subsection{Dynamical formation}

As in our previous work~\cite{DiGiovanni2020a} in this paper we also study the dynamical formation of fermion-boson stars, starting from a generic Gaussian cloud of a bosonic field surrounding an already formed fermionic star. We will focus on the formation of excited stars. As mentioned before, in~\cite{DiGiovanni2020a} we already found the dynamical formation of one excited ($n=1$) fermion-boson star which, in turn, prompted the more detailed investigation we present in the current paper. Here, we will study the dependence of the dynamical formation of excited fermion-boson stars on the initial parameters of the bosonic cloud and of the initial neutron star, populating the stable region of the parameter space. We will limit our study to the $\Lambda=30$ case as a representative choice. 

Our initial data is built by solving the Hamiltonian and momentum constraints. It describes a fermionic star of central density $\rho_c$ surrounded by a bosonic cloud with a Gaussian radial distribution of the form
\begin{eqnarray}
\displaystyle{\phi(r,t) = A_{0} e^{-\frac{r^2}{\sigma^2}} e^{-i\omega t}}\,.
\end{eqnarray}
The freely specifiable parameters $A_{0}$ and $\sigma$ are the amplitude and width of the Gaussian cloud, and $\omega$ is the initial frequency of the scalar field, that we set to one for all models considered. The central density $\rho_c$ is the corresponding free parameter for the fermionic piece. We address the reader to~\cite{DiGiovanni2020a} for further details on the initial-data construction. 

\begin{table*}[t!]
\caption{Representative equlibrium (static) models of excited (one-node) fermion-boson stars. From left to right the columns report the model name, its stability properties, the value of the self-interaction parameter $\Lambda$, the central value of the fluid density $\rho_c$ and of the scalar field $\phi_c$, the field frequency obtained with the shooting method $\omega_{\rm{shoot}}$, the rescaled frequency $\omega$, the total mass $M_{\rm T}$, the ratio of bosons to fermions $N_{\rm B}\mu/N_{\rm F}$, the number of bosons $N_{\rm B}$, the radius containing $99\%$ of bosonic and fermionic particles, $R_{\rm B}$ and $R_{\rm F}$, and the radius containing $95\%$ of the total mass of the star $R_{\rm T}$. All radii are evaluated using Schwarzschild coordinates.}
\centering 
\begin{tabular}{c  c | c c  c c c | c c c c c c}
\hline
\hline   
       
Model & Branch & $\Lambda$ & $\rho_{c}/\mu^2$  & $\phi_c$ & $\omega_{\rm{shoot}}/\mu$ & $\omega/\mu$ & $M_{\rm T}\mu$ & $N_{\rm B}\mu/N_{\rm F}$ & $N_{\rm B}\mu^2$ & $R_{\rm B}\mu$ & $R_{\rm F}\mu$ & $R_{\rm T}\mu$ \\ [0.6ex]
\hline
MS1 & stable & -30 & 0.0011 & 0.017 & 1.2284 & 0.8653 & 1.2743 & 0.0541 & 0.0697 & 13.67 & 9.28 & 8.78  \\
MS2 & stable & -30 & 0.0014 & 0.020 & 1.2717 & 0.8323 & 1.3864 & 0.0490 & 0.0696 & 12.06 & 8.85 & 8.31  \\
MS3 & unstable & -30 & 0.0035 & 0.020 & 1.5427 & 0.6821 & 1.6151 & 0.0138 & 0.0241 & 7.65 & 7.04 & 6.47  \\
MS4 & stable & 0 & 0.0015 & 0.019 & 1.2938 & 0.8267 & 1.4206 & 0.0466 & 0.0681 & 11.90 &  8.72 & 8.18  \\
MS5 & unstable & 0 & 0.0010 & 0.024 & 1.2253 & 0.8885 & 1.1776 & 0.1772 & 0.1873 & 15.28 & 9.28 & 8.78  \\
MS6 & unstable & 0 & 0.0035 & 0.035 & 1.2284 & 0.8653 & 1.2743 & 0.0541 & 0.0697 & 13.67 & 9.19 & 9.31  \\
MS7 & stable & 30 & 0.0020 & 0.032 & 1.4068 & 0.8142 & 1.4403 & 0.1443 & 0.1959 & 11.18 & 7.97 & 7.62  \\
MS8 & unstable & 30 & 0.0017 & 0.033 & 1.3731 & 0.8456 & 1.3651 & 0.2433 & 0.2859 & 12.61 & 8.11 & 8.19  \\
MS9 & unstable & 30 & 0.0025 & 0.045 & 1.5365 & 0.8164 & 1.3958 & 0.3338 & 0.3745 & 10.90 & 7.24 & 7.33  \\
\hline
\hline
\end{tabular}
\label{table:results}
\end{table*}

%%%%%%%%%%%%%%%%%%%%%%%%%%%%%%%%%%%%%%%%%%%%%%%%%%%%
\section{Numerical framework} 
\label{sec:numerics}
%%%%%%%%%%%%%%%%%%%%%%%%%%%%%%%%%%%%%%%%%%%%%%%%%%%%

Both, to study the stability of the equilibrium models as well as their dynamical formation, we resort to numerical-relativity simulations of the Einstein-Klein-Gordon-Euler system, as in~\cite{DiGiovanni2020a}. The numerical evolutions are performed with the numerical-relativity code originally developed by~\cite{Montero:2012yr} and subsequently upgraded to take into account the complex scalar-field equations in~\cite{escorihuela2017quasistationary}. The code employs a second-order Partially Implicit Runge-Kutta method developed by~\cite{Isabel:2012arx, Casas:2014} to evaluate the time update of the evolved quantities. This scheme can handle potential numerical instabilities arising from singular terms appearing in the equations due to our choice of curvilinear coordinates. This computational infrastructure has been extensively tested and used by our group in previous studies of fundamental bosonic fields in strong-gravity spacetimes (see e.g.~\cite{Sanchis-Gual:2015bh,Sanchis-Gual:2015sms,Sanchis-Gual:2015lje,sanchis2017numerical,di2018dynamical,DiGiovanni2020a}). 

To build the initial data we use Schwarzschild coordinates and an equally spaced linear grid, while we use isotropic coordinates and a logarithmic grid in the evolution code. The logarithmic grid allows us to place the outer boundary sufficiently far from the origin and perform long-term stable evolutions.  For the simulations reported in this work we employ a minimum radial resolution of $\Delta_{r} = 0.0125$ with a Courant factor $\Delta_{t} = 0.3 \Delta_{r}$. The inner boundary is set at $r_{\rm{min}}=\Delta r/2$ and the outer boundary is at $r_{\rm{max}}=6000$. We employ 4th-order Kreiss-Oliger numerical dissipation terms to damp spurious high-frequency numerical noise. All advection terms (such as $\beta^{r}\partial_{r}f$) are treated with an upwind scheme. At the outer boundary we impose radiative boundary conditions. The interested reader is addressed to~\cite{Sanchis-Gual:2015sms,DiGiovanni2020a} for further details. We plan to release soon a public version of the code we developed to construct the equilibrium configurations of fermion-boson stars.

 %%%%%%%%%%%%%%%%
\section{Results}
\label{results}
%%%%%%%%%%%%%%%%

\begin{figure*}[t!]
\centering
\hspace{-0.1cm}
\includegraphics[width=0.33\textwidth]{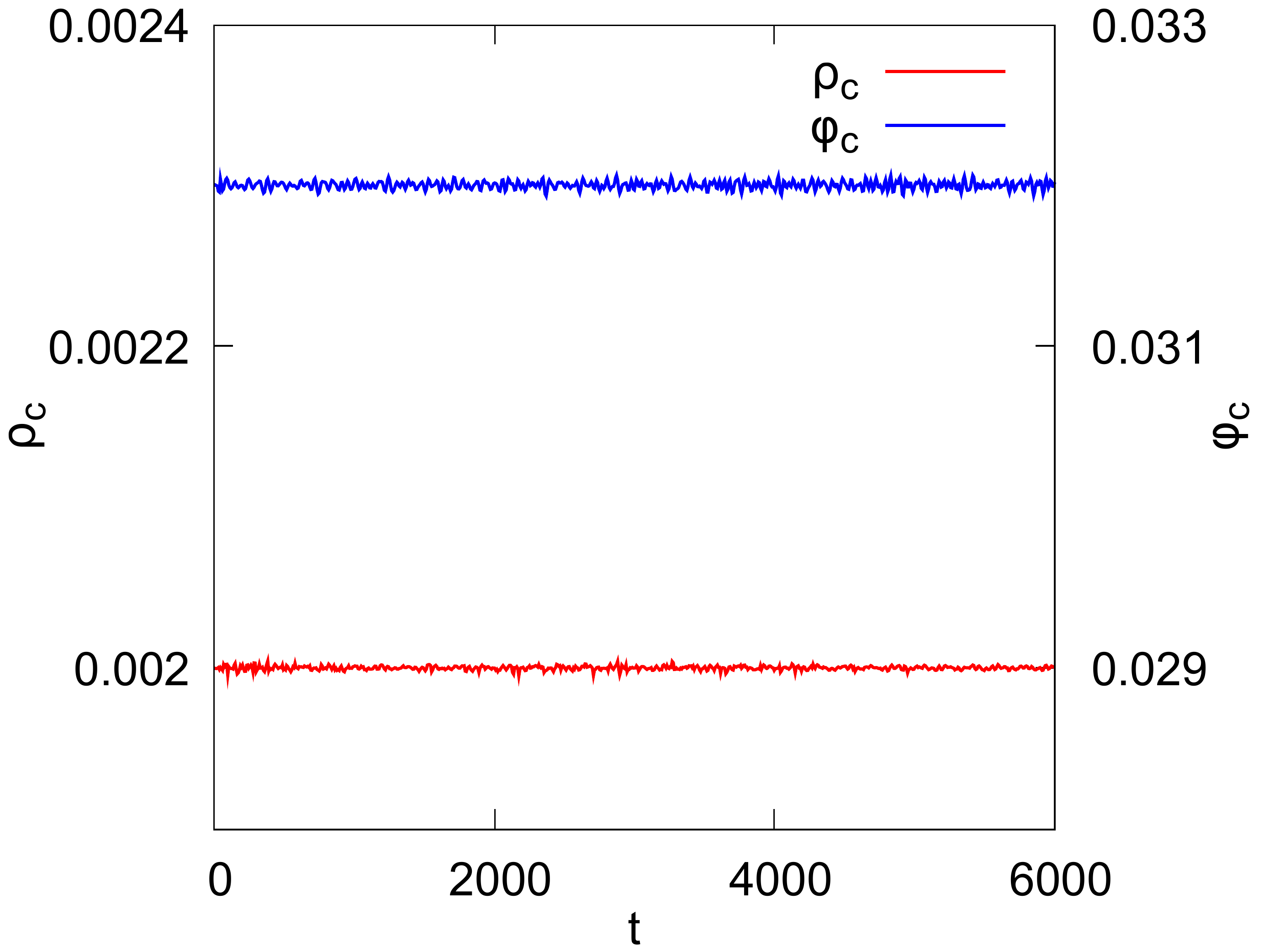} \hspace{-0.2cm}
\includegraphics[width=0.33\textwidth]{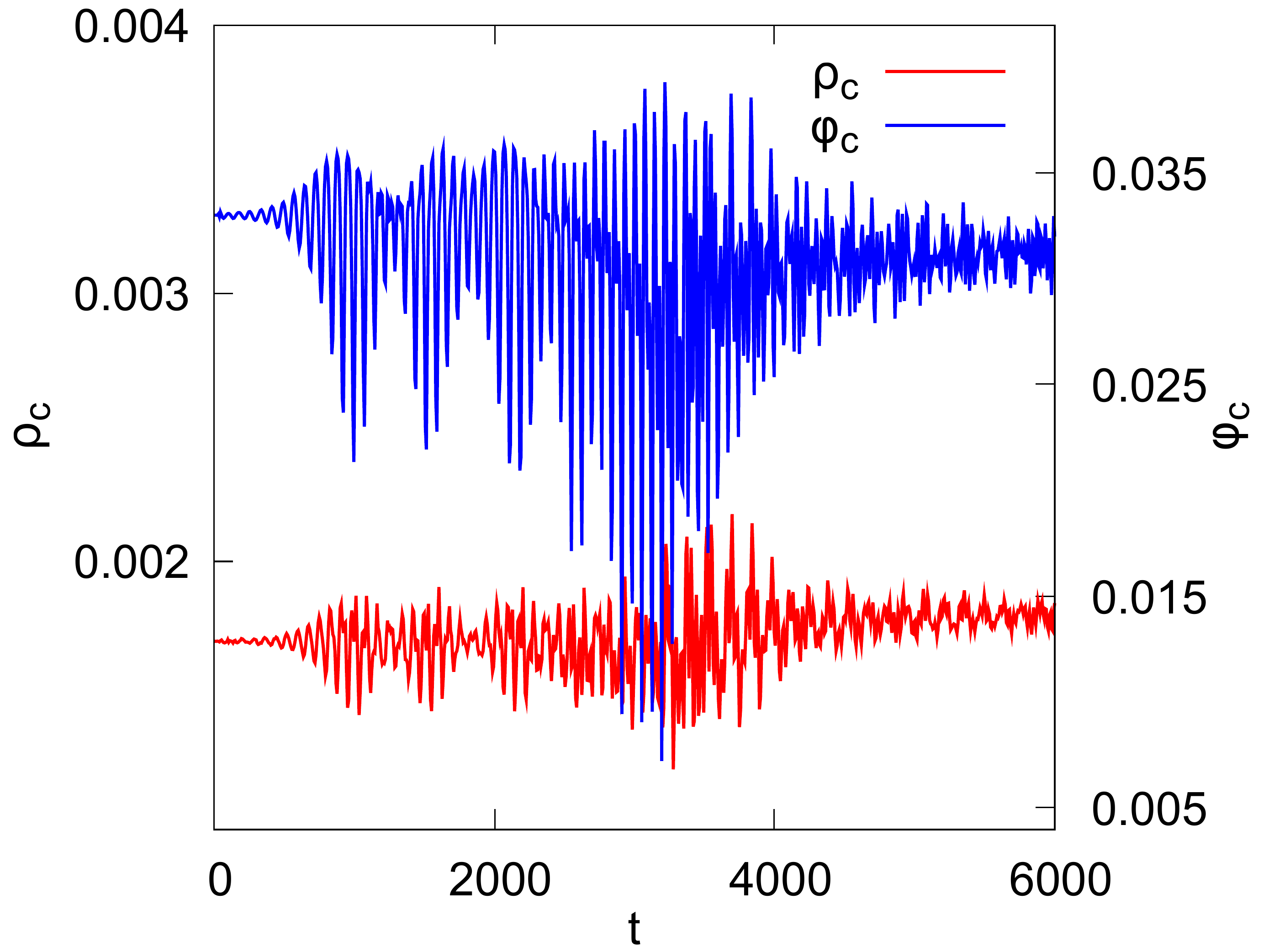} \hspace{-0.07cm}
\includegraphics[width=0.33\textwidth]{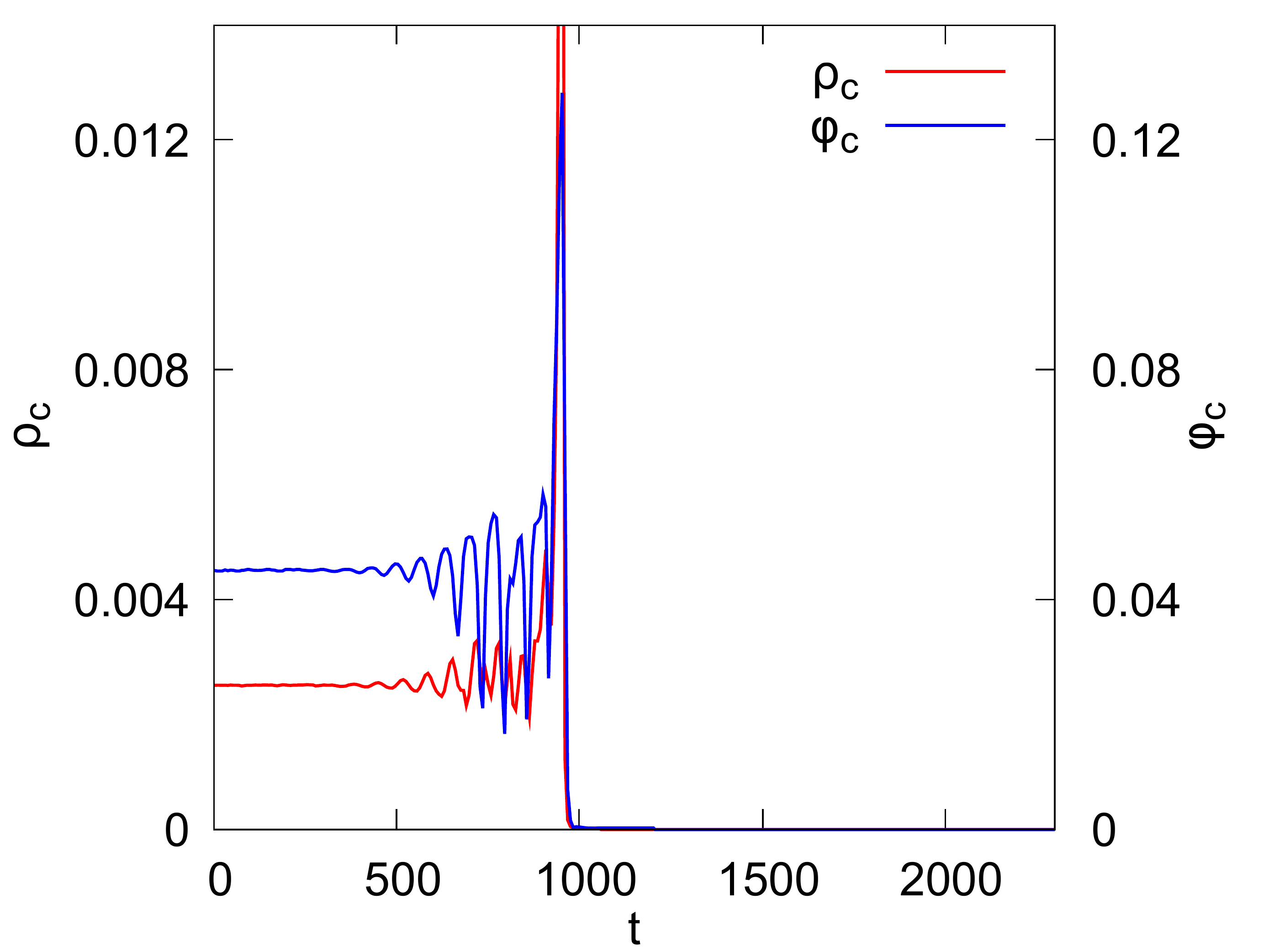} \\ 
\includegraphics[width=0.325\textwidth]{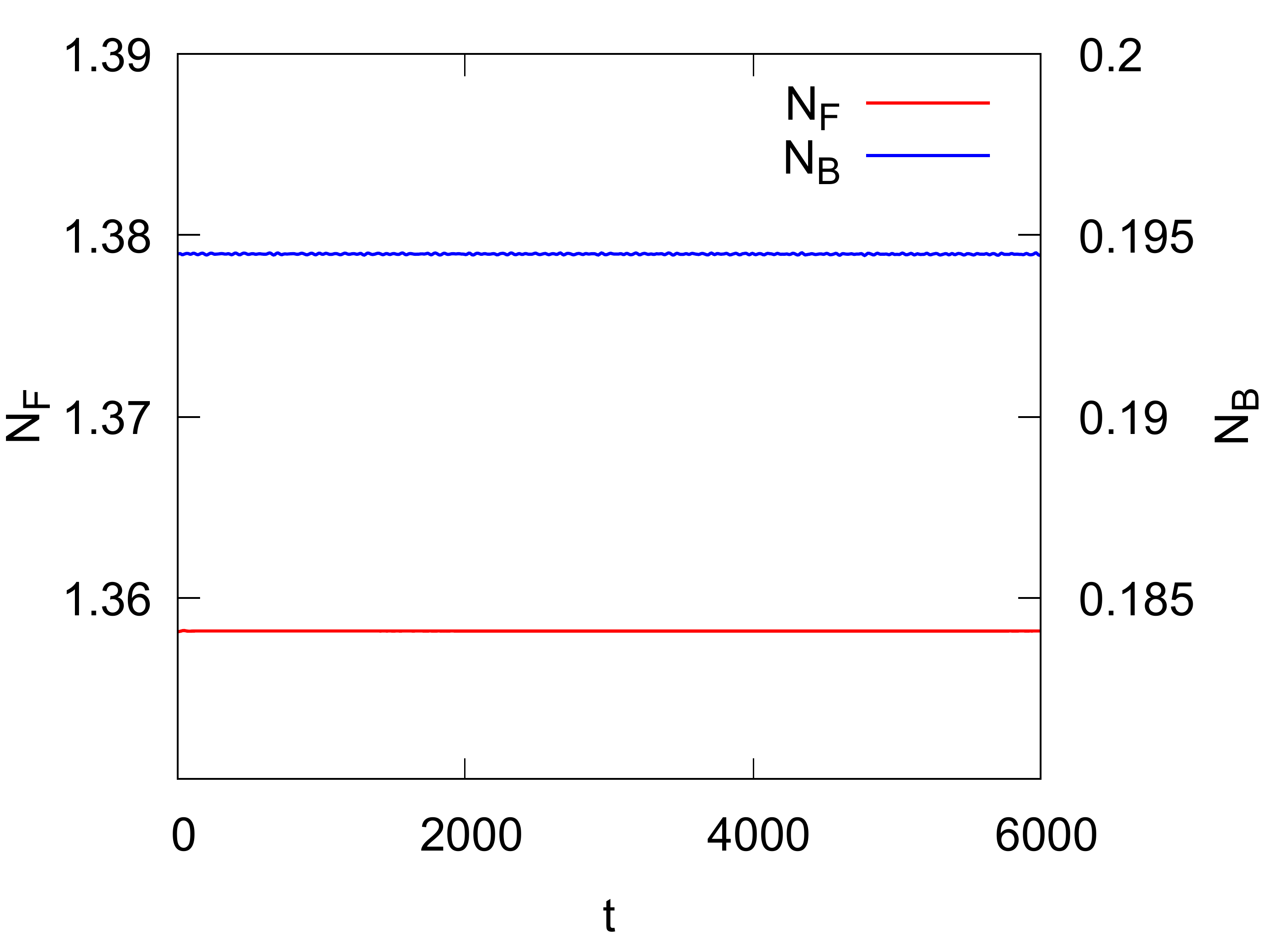} \hspace{-0.13cm}
\includegraphics[width=0.325\textwidth]{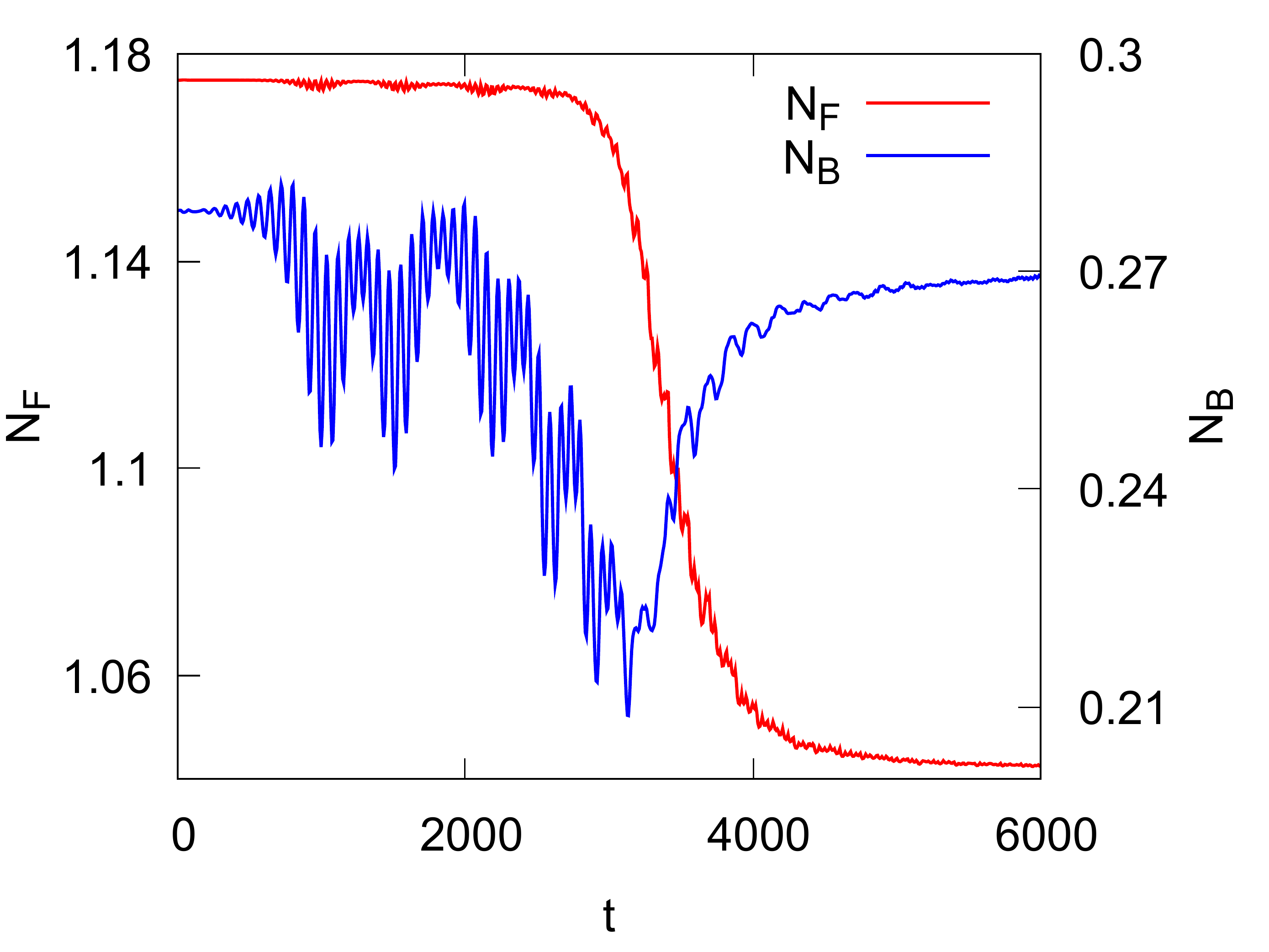} \hspace{-0.07cm}
\includegraphics[width=0.32\textwidth]{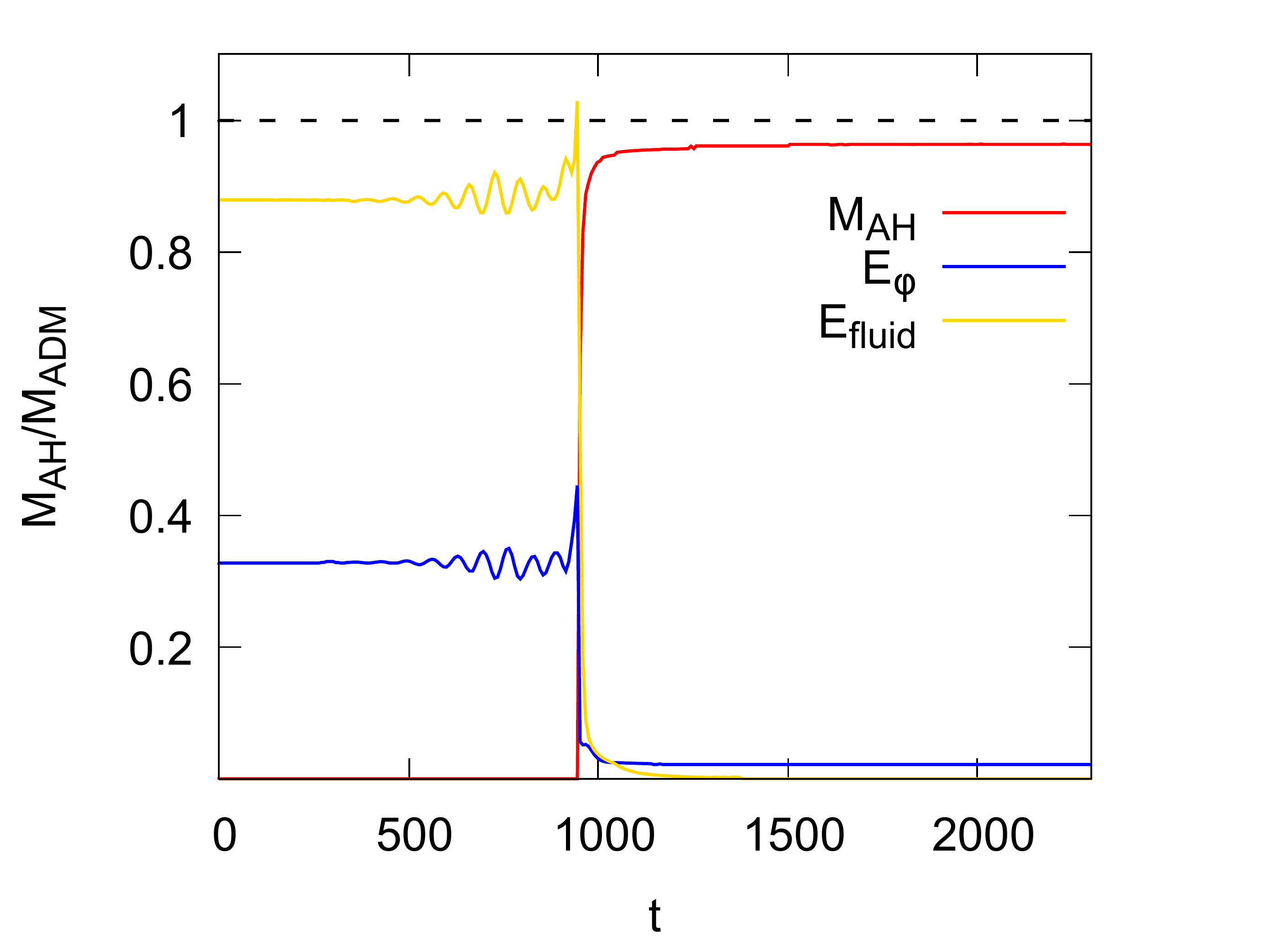}

\caption{Time evolution of representative equilibrium (static) $n=1$ models with self-interaction parameter $\Lambda = 30$. {\it Left column}: stable model MS7. The top row (in all three columns) depicts the central value of the fluid density $\rho_{c}$ and of the scalar field $\phi_{c}$ while the bottom row shows the number of bosons $N_{\rm B}$ and fermions $N_{\rm F}$. {\it Middle column}: migrating (unstable) model MS8. Both rows show the same physical quantities as in the left column. {\it Right column}: collapsing (unstable) model MS9. The bottom row displays the AH mass in units of the ADM mass (red solid line), the volume integrated energy of the scalar field and the fluid, and the time evolution of the ADM mass normalized by its initial value (black dashed line).}
\label{fig:evolutions}
\end{figure*} 

To determine the stability lines in Fig.~\ref{fig:existence_plot} we evolved numerically about ${\cal O}(400)$ models. We turn now to discuss the results for a few representative solutions of the sample to illustrate all possible fates that are expected in evolutions of excited fermion-boson stars. Table~\ref{table:results} reports the most relevant physical properties for nine specific models. For each value of $\Lambda$, namely -30, 0, and 30, we consider one stable model, one model that migrates to a nodeless (ground state) configuration, and one model that collapses to form a black hole, since those are the three possible outcomes. In Fig.~\ref{fig:evolutions} the time evolution of different physical quantities for the three  models with $\Lambda = 30$ are displayed. For the solutions that do not collapse (i.e.~either stable models -- left column -- or models that migrate to the ground state -- central column) we plot the number of bosonic and fermionic particles,  $N_{\rm B}$ and $N_{\rm F}$, and the central values of the scalar field, $\phi_c$, and of the rest-mass fluid density, $\rho_c$. For the model that collapses to form a black hole (right column) we display the apparent-horizon mass, $M_{\rm{AH}}$, the ADM mass of the system, and $\rho_c$, and $\phi_c$. The expected time evolution of any of these representative models can be immediately recognized in the figure.

\begin{table*}[t!]
\caption{Dynamical formation of stable excited fermion-boson stars. The two vertical lines separate the information about the initial model parameters (left), about physical quantities evaluated at the end of the formation process (center), and about  physical quantities of the corresponding equilibrium configuration (right). All models correspond to $\Lambda = 30$ and $\sigma = 70$. Columns on the left box report the central rest-mass density $\rho_{c}$ and the amplitude of the scalar field Gaussian profile $A_{0}$ at the initial time. Columns on the middle box indicate the number of nodes in the radial profile, $n$, the scalar-field frequencies, $\omega_{n}$ (only the one corresponding to the dominant peak in the Fourier transform is reported), the fermionic energy, $E^{\rm{fluid}}_{30}$ within a sphere of radius $r=30$, the bosonic energy, $E^{\phi}_{30}$, the ratio between number of bosons and fermions, $N^{\rm B}_{30}/N^{\rm F}_{30}$, and the ADM mass $M_{\rm ADM}$. Columns on the right box indicate the central values of the rest-mass density and scalar field amplitude, the frequency $\omega$, the fermionic energy $E^{\rm{fluid}}$, the bosonic energy $E^{\phi}$, the ratio between number of bosons and fermions, $N^{\rm B}/N^{\rm F}$, and the ADM mass $M_{\rm ADM}$ of the corresponding equilibrium configuration (with the same number of nodes $n$ in the central box).}
\centering 
\begin{tabular}{c  c c | c c c c c c| c c c c c c c}
\hline
\hline                  
Model & $\rho_{c}/\mu^2$  & $A_{0}$ & $n$ & $\omega_{n}/\mu$ & $E^{\rm{fluid}}_{30}\mu$ & $E^{\phi}_{30}\mu$ & $N^{B}_{30}\mu/N^{F}_{30}$ & $M_{\rm ADM}$ & $\rho_{c}$ & $\phi_{c}$ & $\omega/\mu$ & $E^{\rm{fluid}}\mu$ & $E^{\phi}\mu$ & $N^{B}\mu/N^{F}$ & $M_{\rm ADM}$ \\ [0.5ex]
\hline
MS10 & 0.0008 & $45 \times 10^{-5}$ & 1 & 0.899 & 1.250 & 0.058 & 0.035 & 1.17 & 0.00090 & 0.012 & 0.894 & 1.244 & 0.062 & 0.048 & 1.182  \\
MS11 & 0.0010 & $37 \times 10^{-5}$ & 1 & 0.870 & 1.415 & 0.050 & 0.035 & 1.28 & 0.00110 & 0.012 & 0.868 & 1.407 & 0.048  & 0.033 & 1.295 \\
MS12  & 0.0010 & $25 \times 10^{-5}$ & 2 & 0.923 & 1.419 & 0.015 & 0.014 & 1.28 & 0.00103 & 0.006 & 0.922 & 1.421 & 0.018 & 0.012 & 1.287  \\
MS13 & 0.0010 & $15 \times 10^{-5}$ & 3 & 0.950 & 1.418  & 0.005 & 0.004 & 1.27 & 0.00102 & 0.003 & 0.957 & 1.421 & 0.004 & 0.003 & 1.275  \\
MS14 & 0.0020 & $30 \times 10^{-5}$ & 1 & 0.760 & 1.880 & 0.037 & 0.019 & 1.57 & 0.00227 & 0.016 & 0.760 & 1.863 & 0.036  & 0.018 & 1.576  \\
MS15 & 0.0020 & $25 \times 10^{-5}$ & 3 & 0.897 & 1.883 & 0.020 & 0.007 & 1.57 & 0.00208 & 0.008 & 0.897 & 1.879 & 0.016 & 0.008 & 1.586 \\
MS16 & 0.0020 & $20 \times 10^{-5}$ & 4 & 0.927 & 1.880 & 0.015 & 0.009 & 1.57 & 0.00210 & 0.007 &  0.929 &1.887 & 0.016 & 0.008 &  1.593  \\

\hline
\hline
\end{tabular}
\label{table:models_formation}
\end{table*}

\begin{figure}[t!]
\centering
\begin{minipage}{1\linewidth}
\includegraphics[width=1.0\textwidth]{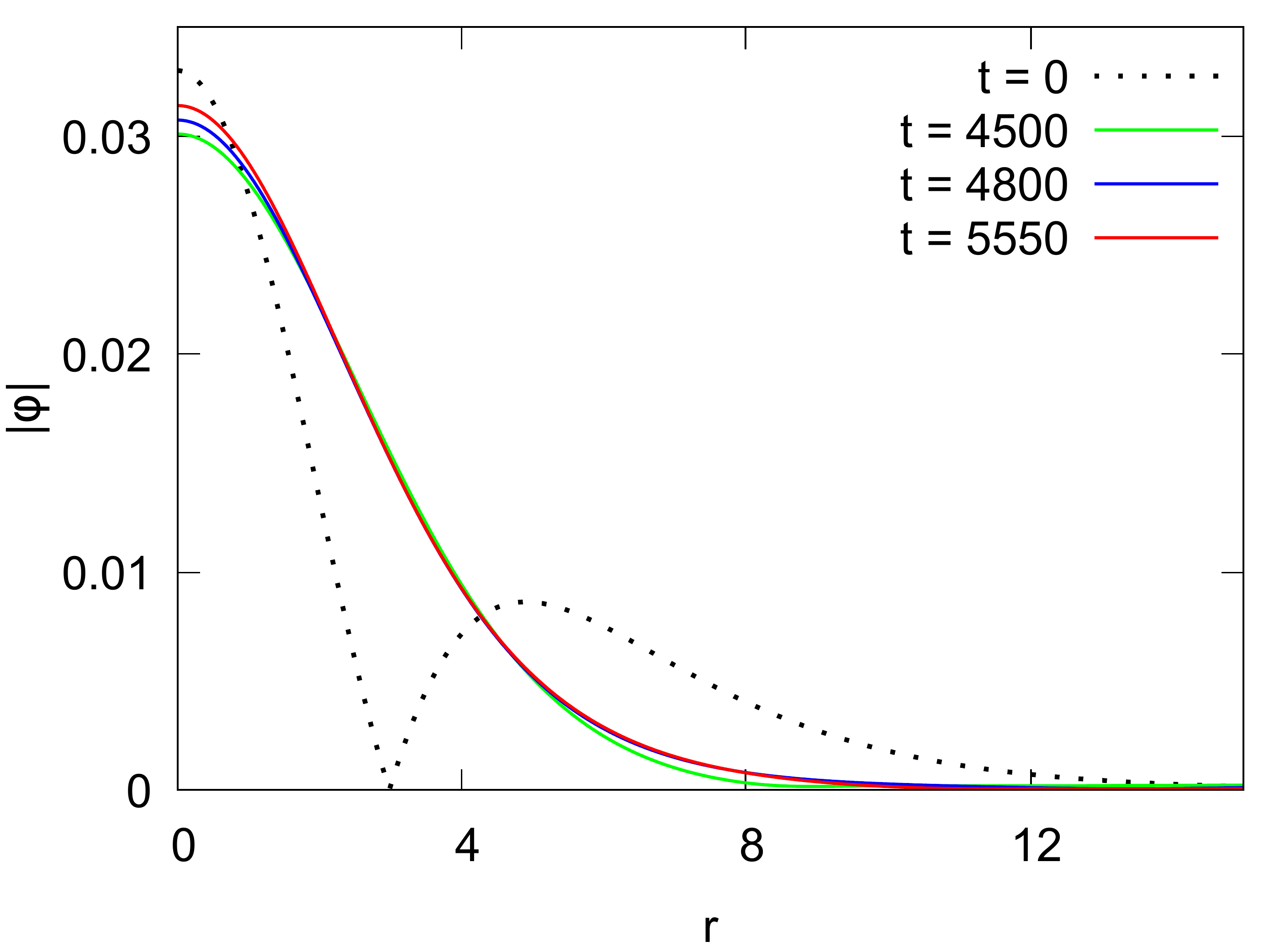} 
\caption{Late-time snapshots of the radial profile of the module of the scalar field $\phi$ for model MS8, compared to the initial state (dashed black line). The model is unstable and migrates to a nodeless configuration.}
\label{fig:phi_migrating}
\end{minipage}
\end{figure}

To better show the results of an excited star  that migrates to the nodeless configuration (model MS8 in Table~\ref{table:results}, also shown in the central panels of Fig.~\ref{fig:evolutions}) we display in Fig.~\ref{fig:phi_migrating} three radial profiles of the scalar field, $\phi(r)$, for late time snapshots, comparing them to the profile of the initial configuration (black dashed line). The evolution clearly exhibits that this model is indeed unstable and migrates to a stable ground-state fermion-boson star where no nodes are visible across the star. The final profiles neatly oscillate around a new stable configuration. 

Once the stability of equilibrium configurations of excited mixed stars has been established, we turn our attention to the dynamical formation scenario. We performed simulations varying the shape of the bosonic cloud, considering different initial fermionic star models. A subset of those models, namely those for $\sigma=70$, are reported in Table~\ref{table:models_formation}. Our simulations show that excited mixed stars with one or even more nodes in the radial profile of $\phi$ can indeed form dynamically from the collapse/accretion of an initial bosonic cloud through the gravitational-cooling mechanism. By keeping fixed the cloud width $\sigma$ we observe that the lower  the amplitude $A_{0}$ of the initial cloud, the lower the final value of $\phi_{c}$ and the more radial nodes appear in the scalar-field profile. This means that excited states with nodes are {\it preferred} final configurations to ground-state mixed stars for some region of the parameter space. %Such configurations can store more energy and scalar-field mass than those at the ground state. 

\begin{figure}[t!]
\centering
\begin{minipage}{1\linewidth}
\includegraphics[width=0.9\textwidth]{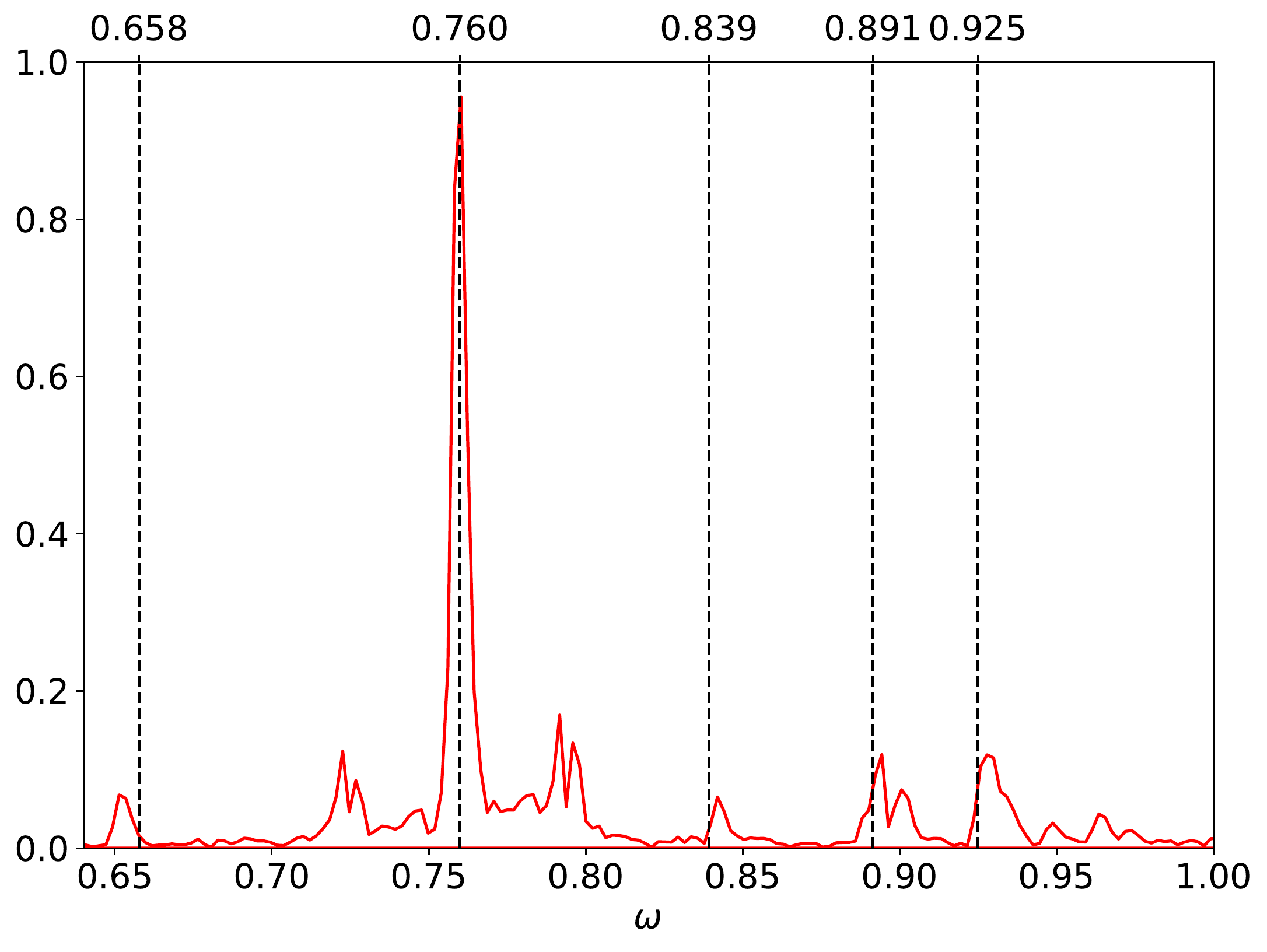} 
\includegraphics[width=0.9\textwidth]{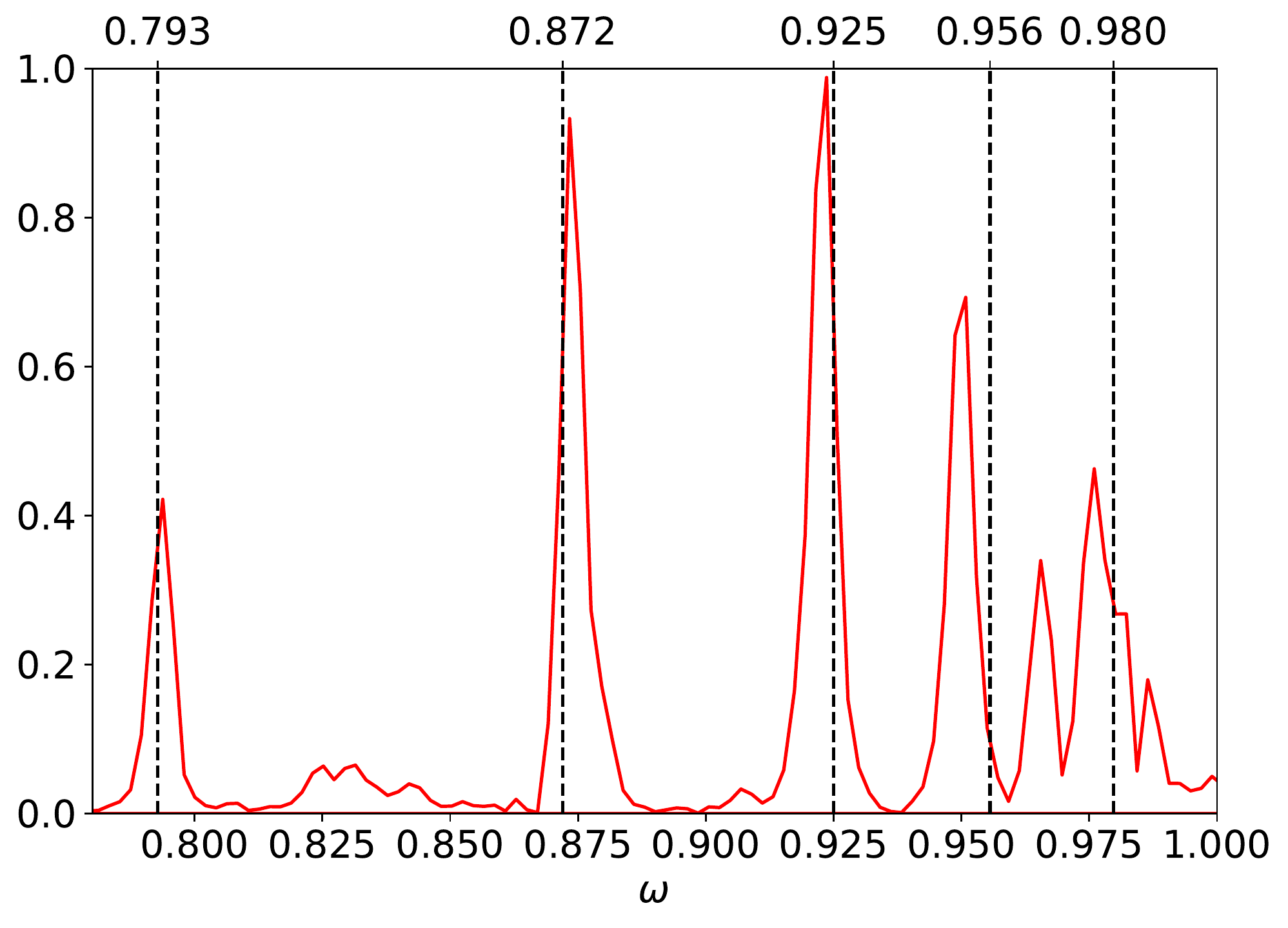} 
\caption{Fourier transform of the time evolution of the scalar field $\phi$ for models MS14 ({top}) and MS12 ({bottom}). From left to right, the vertical dashed lines correspond to the values of the frequencies of the ground state and of the first four excited states of static models similar to the end result of models MS14 and MS15 (the specific values are reported at the top of each panel). An apparent peak corresponding to the $n=1$ excited state can be seen for model MS14 while the coexistence of more than one excited state is visible for MS12. The units of the vertical axes are arbitrary.}
\label{fig:Fourier}
\end{minipage}
\end{figure} 

The final object resulting from the (incomplete) gravitational collapse of a bosonic cloud is always radially perturbed. We notice that some of the objects oscillate between different states with zero, one or more nodes. The region of stability of the excited stars becomes smaller as the number of nodes in the scalar field increases, shrinking towards the neutron star stability lines in Fig.~\ref{fig:existence_plot}. From these findings we hypothesize that if the final configuration resides in a region of the parameter space where several stable excited states exist, the perturbation that the object undergoes due to the gravitational cooling process can cause the migration to a different state of the scalar field. 

\begin{figure}[t!]
\centering
\begin{minipage}{1\linewidth}
\includegraphics[width=0.9\textwidth]{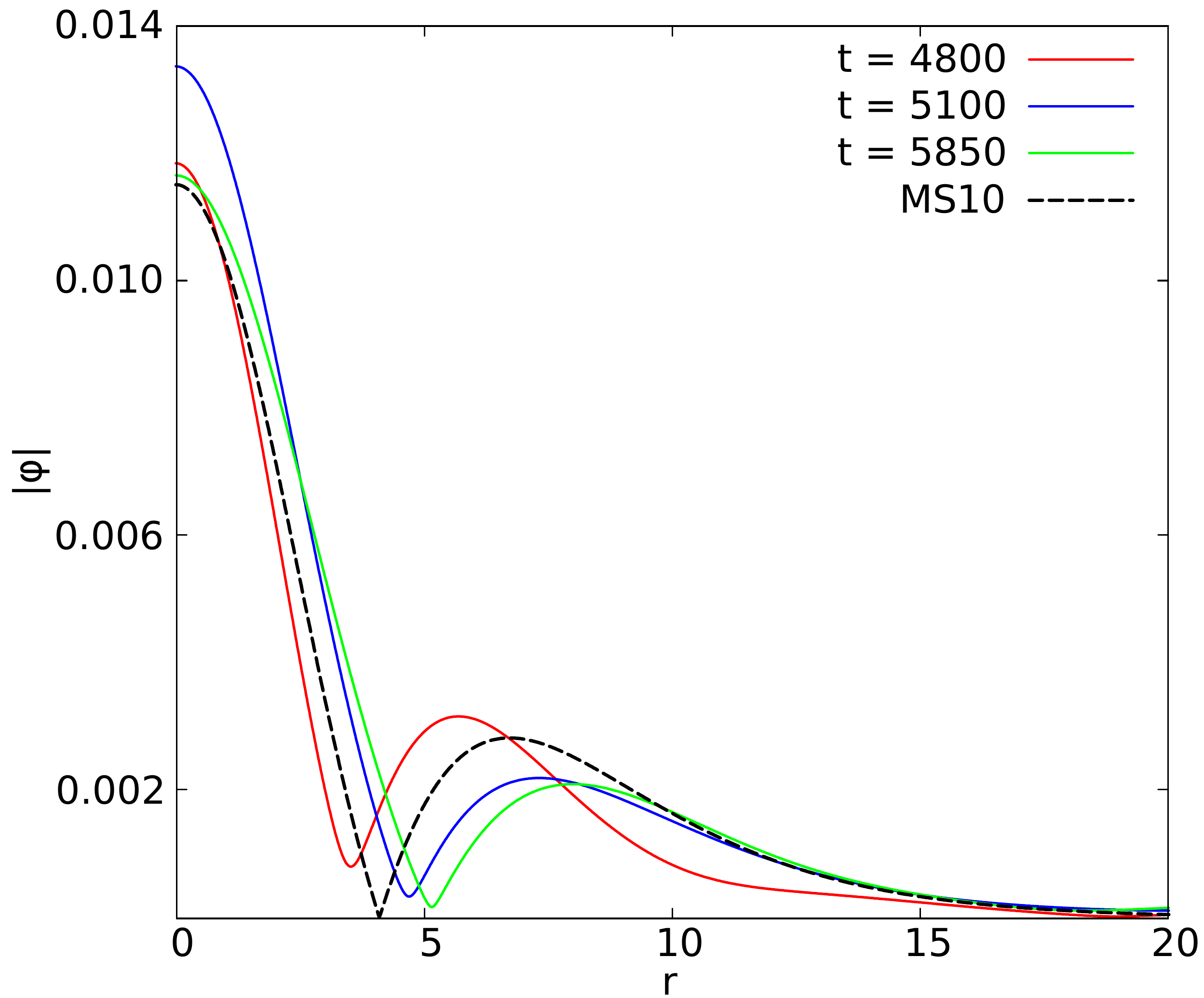} 
\includegraphics[width=0.95\textwidth]{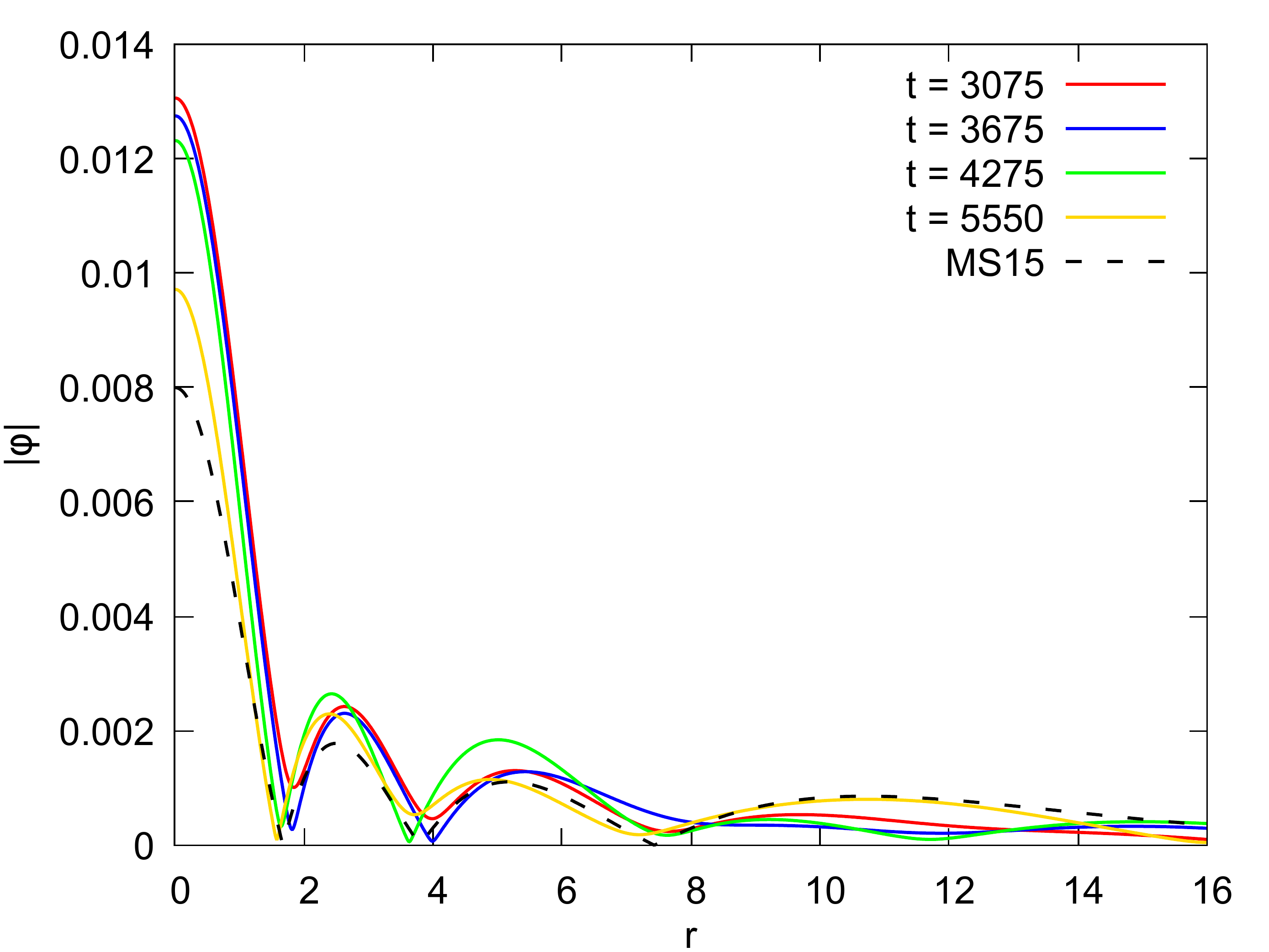} 
\caption{Late-time snapshots of the radial profile of the module of the scalar field $\phi$ for models MS10 (top) and MS15 (bottom). The dashed black lines indicate the profiles of the corresponding static models with similar $\rho_{c}$ and $\phi_{c}$.}
\label{fig:latetime_phi}
\end{minipage}
\end{figure} 

Table~\ref{table:models_formation} reports a summary of the parameters of the initial data that we have evolved and the relevant physical quantities of the final object. Those are compared to the same quantities of a static configuration with similar properties. In particular the table reports the number of nodes, $n$, and the oscillation frequency of the models. The latter is evaluated by means of a Fast-Fourier transform of the scalar field amplitude in the time window $t\in[4000,6000]$. The frequency reported is the one corresponding to the dominant peak in the Fourier transform. As an illustrative  example we show in Fig~\ref{fig:Fourier} the Fourier transform of models MS14 and MS12. In the first case we see  a larger peak corresponding to the dominant frequency of the first excited state, while for the latter the peaks corresponding to the first, second and third excited states are comparable. This means that the final object resulting from the evolution of model MS12 is oscillating between these three configurations. 

Fig.~\ref{fig:latetime_phi} depicts a few late-time snapshots of the radial profiles of the module of the scalar field, $|\phi(r)|$, after the formation process has been completed. We show two different initial data setups, namely MS10 and MS15. We compare the objects formed dynamically with the corresponding static solutions with similar physical properties (in terms of mass and oscillation frequency) to identify the stars. The dashed black lines in Fig~\ref{fig:latetime_phi} display the radial profiles corresponding to those static solutions. As we can see, model MS15 shows three distinctive nodes in the scalar-field radial profile as a result of its dynamical formation.

%%%%%%%%%%%%%%%%%%%%%%%%%%%%%%%%%%%%%%%%%%%%%%%%%%%%%%%%%%%%%
\section{Discussion}
\label{sec:conclusions}

In this paper we have studied the nonlinear stability of {\it excited} fermion-boson stars in spherical symmetry, i.e.~models for which the radial profile of the scalar field shows at least one node across the star. This investigation has extended previous results concerning the dynamical properties of fermion-boson stars (see 
e.g.~\cite{PhysRevD.87.084040,PhysRevD.102.064038,DiGiovanni2020a}) by considering for the first time a complex scalar field in an excited state. We have constructed (hundreds of) equilibrium (static) configurations of excited fermion-boson stars with and without a quartic self-interaction term in the potential, studying their evolution in order to identify possible regions of stability in the parameter space. For purely spherically symmetric boson stars, excited configurations are known to be generically unstable~\cite{Lee:1988av,Balakrishna:1997ej}.  Through numerical-relativity simulations of the Einstein-Klein-Gordon-Euler system we have shown that the presence of enough fermionic matter, in the form of a neutron star, can stabilise the (otherwise unstable) excited scalar field. Such a cooperative mechanism between the fermionic and bosonic constituents of a compound star allows for the existence of a stable region in the parameter space of solutions. Our results have thus confirmed that excited mixed stars with one node can indeed be stable, first reported in~\cite{DiGiovanni2020a}, and their existence plots  have been studied in detail here. Moreover, we have also observed that fermion-boson stars with even more than one node in the radial profile of the scalar field can also be stable. 

In addition to building static models of excited fermion-boson stars we have also analyzed their dynamical formation. To do so we have constructed constraint-satisfying initial data describing a neutron star, modelled by a zero-temperature polytropic EoS, surrounded by an accreting Gaussian cloud of a massive, complex scalar field. These initial data have been evolved to study the potential formation of excited compound stars through the gravitational cooling mechanism. Our results have shown that, depending on the initial parameters of the cloud, different final states can be reached, corresponding to fermion-boson star models either in the ground-state or in an excited state. In some cases, the perturbed final object resulting from the formation oscillates between different scalar-field states with zero, one or more nodes. This outcome is possible because there exist regions in the parameter space of ($\phi_{c}, \rho_{c}$) that can be populated by more than one stable configuration corresponding to states of the scalar field with different number of nodes. The final object that resides in such a region can therefore migrate from one configuration to another due to the perturbation given by the gravitational cooling mechanism, which can yield a positive or negative contribution to the energy stored in the scalar field. Such an outcome is not possible in the case of boson stars since all the excited-state solutions are unstable. Tentatively, this process might be compared to the excitation of the Hydrogen atom, occurring when the electron gains (or loses) the sufficient amount of energy to move from one shell to another.

Our findings confirm that the instability of excited boson stars can be quenched by considering the superposition of two stars, one being stable, that only interact through gravity, irrespective of the type of matter of the stable star. This cooperative stabilization mechanism has already been shown to operate against different instabilities in boson stars~\cite{Bernal:2009zy,Alcubierre:2018ahf,Guzman:2019gqc,Jaramillo:2020rsv,guzman2021stability,sanchis2021multi}. The presence of a second (or a third) star strengthens the stability properties of unstable compact objects that lay in their linearly stable branch. Therefore, those unstable configurations, when combined with other stars, give rise to new mixed objects that can modify the stability properties of both constituents.

%%%%%%%%%%%%%%%%%%%%%%%%%%%%%%%%%%%%%%%%%%%%%%%%%%%%%%%%%%%%%
%%%%%%%%%%%%%%%%%%%%%%%%%%%%%%%%%%%%%%%%%%%%%%%%%

\acknowledgments
We thank Eugen Radu, Carlos Herdeiro and Alexandre Pombo for useful 
suggestions. 
This work was supported by the Spanish Agencia Estatal de Investigaci\'on (grant 
PGC2018-095984-B-I00), by the Generalitat Valenciana (PROMETEO/2019/071 and 
GRISOLIAP/2019/029), by the European Union’s Horizon 2020 RISE programme 
H2020-MSCA-RISE-2017 Grant No.~FunFiCO-777740, by DGAPA-UNAM through grant 
No.~IN105920, by 
CONACyT Ciencia de Frontera Projects 
No. 376127 ``Sombras, lentes y ondas gravitatorias generadas por objetos
compactos astrof\'\i sicos", and No. 304001 ''Estudio de campos escalares con 
aplicaciones en cosmolog\'ia y astrof\'isica", by the Funda\c c\~ao para a 
Ci\^encia e a Tecnologia (FCT) projects PTDC/FIS-OUT/28407/2017, PTDC/FIS-AST/3041/2020 and UID/FIS/00099/2020 (CENTRA), and CERN/FIS-PAR/0027/2019. SF gratefully acknowledges support by the Erasmus+ International Credit Mobility Program KA-107 for an academic stay at the University of Valencia.

%%%%%%%%%%%%%%%%%%%%%%%%%%%%%%%%%%%%%%%%%%%%%%%

\bigskip

%\newpage

\bibliography{num-rel2}

\end{document}